\newif\ifsubmode 
\submodefalse 

\ifsubmode 
  \documentclass[12pt,preprint]{aastex}  

 \else 
   \documentclass{emulateapj}   
   \usepackage{lscape} 
   \slugcomment{Accepted for publication in ApJ} 
 \fi 

\shortauthors{Fassnacht et al.}
\shorttitle{{\em Chandra} observations of lens groups}

\begin{document}


\newcommand{\kms}{km\ s$^{-1}$}
\newcommand{\kmsmpc}{km\ s$^{-1}$\ Mpc$^{-1}$}

\title{The X-ray Properties of Moderate-redshift Galaxy Groups Selected
by Association with Gravitational Lenses}

 \author{
  C. D. Fassnacht,
  D. D. Kocevski,
  M. W. Auger,
  L. M. Lubin,
  J. L. Neureuther
}
\affil{
  Department of Physics, 
  University of California, Davis, 
  1 Shields Avenue, 
  Davis, CA 95616
}
\email{
  fassnacht@physics.ucdavis.edu
}

\author{
  T. E. Jeltema
}
\affil{
   UCO/Lick Observatories, 
   1156 High St., 
   Santa Cruz, CA 95064
}

\author{
  J. S. Mulchaey
}
\affil{
  Observatories of the Carnegie Institute of Washington,
  813 Santa Barbara St.,
  Pasadena, CA 91101 
}

\author{
  J. P. McKean
}
\affil{
  Max-Planck-Institut f\"{u}r Radioastronomie, 
  Auf dem H\"{u}gel 69, 
  D-53121 Bonn, Germany
}

\begin{abstract}

We present results from a systematic investigation of the X-ray
properties of a sample of moderate redshift ($0.3<z<0.6$) galaxy groups.
These groups were selected not by traditional X-ray or optical search
methods, but rather by an association, either physical or along the
line of sight, with a strong gravitational lens.  We calculate the
properties of seven galaxy groups in the fields of six lens systems.
Diffuse X-ray emission from the intragroup medium is detected in four
of the groups.  All of the detected groups have X-ray luminosities
greater than $10^{42} h^{-2}$ erg s$^{-1}$, and lie on the
$L_X$--$\sigma_v$ relations defined by local groups and clusters.  The
upper limits for the non-detections are also consistent with the local
$L_X$--$\sigma_v$ relationships.  Although the sample size is small
and deeper optical and X-ray data are needed, these results suggest
that lens-selected groups are similar to X-ray selected samples and,
thus, are more massive than the typical poor-group environments of
local galaxies.

\end{abstract}

\keywords{
   galaxies: clusters: general ---
   gravitational lensing ---
   X-rays: galaxies: clusters
}

\section{Introduction}

To have a full picture of galaxy evolution and structure formation in
the Universe, it is crucial to understand the properties of galaxy
groups.  Most galaxies in the local Universe reside in galaxy groups
\citep[e.g.,][]{turnergott,gellerhuchra,eke04}.
In addition, groups are vitally important in driving changes in star
formation rates and galaxy morphologies since $z \sim 1$, because the
low velocity dispersions and high density of groups make them likely
locations for interactions and mergers
\citep[e.g.,][]{aarsethfall,barnes85,merritt85}.  Furthermore,
indications are that the dark matter distributions in groups represent
a transition between the dark-matter dominated profiles seen on cluster
scales and galaxy-sized halos that are strongly affected by baryon
cooling \citep[e.g.,][]{oguri05}.  In this paper we examine the properties
of moderate-redshift groups.

Groups in the local Universe have been well studied 
\citep[e.g.,][]{zm98,mz98,gemsxray,rasmussen06}.
In a systematic survey of 60 groups in the Group Evolution
Multiwavelength Study (GEMS), \citet{gemsxray} find that groups in the
local Universe obey similar $L_X$--$\sigma_v$ and $L_X$--$T$ scaling
relations to more massive clusters, although a more recent analysis
\citep{gems_refit} finds a steepening in the $L_X$--$T$ slope for
these groups.  There is also a large non-statistical scatter of a
factor of 30 in X-ray luminosity and a factor of 3--4 in X-ray
temperature for this group sample.  This scatter includes a class of
spiral-rich groups with little or no emission, as well as some groups
with high X-ray luminosities but low velocity dispersions. Possible
explanations for these outliers include unrelaxed dynamical states,
uncertain velocity dispersion measures from small numbers of redshifts
\citep[typically fewer than 10, see also][]{zm98}, and point source
contamination of the X-ray flux \citep{helsdon05}. \citet{gemsxray}
also find a strong correlation between the detection of diffuse
emission and the presence of a centrally-located, dominant early-type
galaxy, as well as an anti-correlation between spiral fraction and gas
temperature.

Additional outliers from the canonical scaling relations are found by
\citet{rasmussen06}, who are studying a redshift-selected,
statistically-unbiased sample of galaxy groups at $z \approx 0.06$
with deep X-ray data. The first galaxy groups detected from this
survey are X-ray faint ($\sim 5 \times 10^{40}~h^{-2}$ erg s$^{-1}$), are
underluminous for their measured velocity dispersions, and do
not host a dominant, central early-type galaxy, suggesting that they
are collapsing for the first time. Optically-selected groups such as
these represent a different, perhaps more common, class of groups than
those detected through their X-ray emission.

Detailed studies of groups at moderate-to-high redshifts have been
limited because groups are difficult to find, given their modest
galaxy overdensities and X-ray luminosities.  On the optical side, the
situation has been alleviated somewhat by large redshift surveys such
as the Sloan Digital Sky Survey \citep[SDSS;][]{sdss}, the Canadian
Network for Observational Cosmology Field Galaxy Survey
\citep[CNOC2;][]{cnoc2}, and the DEEP2 survey \citep{deep2}, but many
of the group candidates found in these surveys have only 3-4 members
\citep[e.g.,][]{cnoc2groups1,deep2groups} and may not represent
physically bound structures.  Furthermore, the intensive spectroscopic
followup required to confirm each group candidate and characterize its
properties has so far limited the size of well-studied group samples
from these surveys \citep[e.g.,][]{wilmangrp1,balogh07,fang07}.

Intermediate-redshift ($0.2 < z < 0.6$) groups selected on the basis
of their X-ray emission are also being studied 
\citep{xmmlss_groups,xmmlss_erratum,mulchaey96,jeltema06,jeltema07}.
Results for these surveys suggest that these groups follow the
low-redshift scaling relations between luminosity, temperature, and
velocity dispersion. Many of these systems have high fractions of
early-type galaxies, suggesting that this population is already in
place by $z \sim 0.5$ \citep{mulchaey96,jeltema07}. However,
there are clear indications of dynamical evolution, including many
groups with brightest group galaxies that show multiple components,
dominant early-type galaxies that are not centered on the diffuse
emission, or no dominant early-type galaxies at all. While similar
systems have been identified in low-redshift samples (see above), the
luminosities and temperatures of the moderate-redshift examples are
significantly higher, implying a group downsizing where more massive
groups are still in the process of collapsing and virializing at these
redshifts. The effects of evolution appear to continue to higher
redshifts where studies of some optically-selected groups suggest that
they are substantially and systematically underluminous, relative to
their local counterparts, for a given velocity dispersion
\citep{fang07}.  Note however, that the velocity dispersions in
\citet{fang07} are all based on a small number of measured redshifts
(ranging from 3 to 6 redshifts per group) and thus are highly uncertain
and may be significantly overestimated.

Significant differences between the properties of X-ray and
optically-selected groups have been noted at both low and intermediate
redshift \citep[e.g.,][]{rasmussen06,fang07,rykoff08}, 
suggesting that each method is selecting a distinct class of
groups. To avoid these biases, we are conducting a survey of
moderate-redshift groups that have been selected via a non-traditional
technique, namely through their association with strong gravitational
lenses.  There is growing evidence that strong gravitational lenses,
i.e., those forming multiple images of the background object, are
typically located in groups of galaxies at intermediate redshift
\citep[e.g.,][]{kundic1115,kundic1422,tonrygroups,momcheva,1608groupdisc,auger_groups1,auger1520}. 
Thus, strong lenses can be used to identify and study the properties
of distant groups, selected in a manner that is completely independent
from alternative techniques such as deep X-ray integrations or
color-magnitude diagrams.  The lensing probability depends only on the
projected mass distribution and does not depend on its kinematics or
on the properties of the galaxy population or intragroup medium
(IGM). Therefore, lens-selected groups provide an excellent sample to
determine the properties of galaxies and hot baryons, and to
understand selection effects (e.g., mass concentration for lensing,
IGM luminosity and temperature for X-ray selection, and the
homogeneity of the galaxy population for red sequence selection) by
contrasting different methods.

It is not unexpected that the lensing galaxies in strong lens systems
should reside in overdense regions of the Universe.  Searches for
gravitational lenses are biased toward high density regions because
(1) higher mass systems have a larger cross-section for lensing and
(2) most lensing galaxies are ellipticals, which are preferentially
found in groups or clusters \citep[e.g.,][]{morphdensity,zm98}.
Theoretical studies have predicted that a significant number of lens
systems should be associated with groups or clusters, albeit with a
large spread of values
\citep[]{keeton00,holderschechter,oguri_etal05}.  Unbiased photometric
surveys of lens fields indicate that lenses lie along overdense lines
of sight
\citep[][Fassnacht et al.\ in prep.]{williams06}, and spectroscopic
observations have confirmed several lens-group associations that can
affect the lensing potential at the level of 5\% or more
\citep[e.g.,][]{kundic1115,kundic1422,tonrygroups,0712group,momcheva,1608groupdisc,auger_groups1,auger1520}.

In this paper, we present the results from multi-object spectroscopic
and deep {\em Chandra} observations of seven groups detected through their
association with gravitational lenses.  We assume a cosmological
model with $(\Omega_{\rm M},\Omega_\Lambda) = (0.3,0.7)$.  We will
use $h = H_0 / 100$\kmsmpc\ to represent the Hubble Constant when
we do not have to choose a value.  When we do have to assign a
value to the Hubble Constant, we use $h = 0.7$.

\section{The Sample}

The systems analyzed for this paper are all strong lenses for which we
have either obtained new observations with {\em Chandra} or for which
data are available in the {\em Chandra} archive.  Furthermore, each
system had to show evidence of an associated galaxy group, obtained
through spectroscopic surveys of the galaxies surrounding the main
lens system.  The targeted systems are described briefly below.

\subsection{CLASS B0712+472}

This four-image lens system (hereafter B0712) was discovered by
\citet{0712disc} as part of the Cosmic Lens All-sky Survey
\citep[CLASS;][]{class1,class2}.  The lensing galaxy is at a redshift
of $z_{\rm lens} = 0.406$ \citep{zclass}, while the lensed
source is at a redshift of 1.34 \citep{0712disc}.  A spectroscopic
survey discovered a group in the foreground of the lens, with 10
confirmed members and a mean redshift of $z = 0.29$ \citep{0712group}.
Further spectroscopic followup, presented in this paper, finds 5 more
members.

\subsection{PG 1115+080}

This lens system (hereafter PG1115), consisting of four lensed images
of a $z_{\rm source} = 1.722$ quasar, was the second lens discovered
\citep{1115disc}.  The lens redshift is $z_{\rm lens} = 0.310$
\citep{kundic1115,tonrygroups}.  The system is especially important
because it is one of only $\sim$10 lenses for which time delays have
been measured \citep{1115delay_schech,1115delay_barkana}.  Early observations
of this system found a likely group of galaxies
centered close to the lens system \citep{young1115} and found that
two of the potential group members had redshifts of $\sim$0.3
\citep{henry1115}, suggesting that the lensing galaxy was a member
of a small group.  The group membership was expanded by
\citet{kundic1115} and \citet{tonrygroups} who between them found 5
group members, including the lensing galaxy.  Recent work by
\citet{momcheva} has extended the number of spectroscopically 
confirmed members to 13.

\subsection{JVAS B1422+231}

This is a four-image lens system (hereafter B1422) discovered by
\citet{1422disc} as part as the Jodrell-VLA Astrometric Survey
\citep{jvas1,jvas2,jvas3}.  The background source is at a redshift 
of $z_{\rm source} = 3.62$, while the lensing galaxy is at $z_{\rm
lens} = 0.647$ \citep{hammer1422}.  Early models of the system
\citep[e.g.,][]{hogg1422}  suggested that external mass was
necessary, and subsequent spectroscopy \citep{kundic1422,tonrygroups}
revealed a group at the redshift of the lens.  The work of
\citet{momcheva} has brought the number of spectroscopically confirmed
group members to 16.

\subsection{CLASS B1600+434}

This two-image system (hereafter B1600) was one of the first two
lenses discovered
\citep{1600disc} in the CLASS survey.  The system redshifts are
$z_{\rm source} = $1.59 \citep{1600disc} and $z_{\rm lens} = $0.414
\citep{zclass}.  This is another time delay system, with delays
measured by \citet{1600delay_radio} at radio wavelengths and
\citet{1600delay_opt} at optical wavelengths.  A spectroscopic
survey has discovered a small group with 7 confirmed members that is
associated with the lens \citep{auger_groups1}.  

\subsection{CLASS B1608+656}

This four-image system (hereafter B1608) was the second lens discovered
\citep{stm1608} at the beginning of the CLASS survey.  The lens
redshift is $z_{\rm lens} = 0.630$ \citep{stm1608}, while the source
redshift is $z_{\rm source} = 1.39$ \citep{zs1608}.  All three
independent time delays in this system have been measured 
\citep{1608delays1,1608delays2}.  An extensive spectroscopic survey of the field
has revealed four candidate galaxy groups along the line of sight to
the lens, with mean redshifts of 0.26, 0.43, 0.51, and 0.63
\citep{1608groupdisc}.  In this paper, we present new spectroscopy of
the field and updated group velocity dispersions.  As we will
discuss below, we will only concentrate on the properties of two of the
groups in this field: the group that is physically associated with
the lensing galaxy at $z = 0.632$ (hereafter B1608-1) and the group
at $z = 0.426$ 
\citep[hereafter B1608-3, using the notation of][]{1608groupdisc}.

\subsection{CLASS B2108+213}

This two-image system (hereafter B2108) has a lensing galaxy at
$z_{\rm lens} = 0.365$ and has the largest image separation (4\farcs6)
of the CLASS lenses \citep{2108disc}, giving a strong indication that
the lens resides in a group or cluster environment.  Unusually, both
the lensed source and the lensing galaxy are radio loud, with the
lensing galaxy showing both a flat-spectrum core and low surface
brightness lobe extending in roughly a east-west to SE-NW direction
\citep{2108disc,more2108}.  The source redshift has not yet been
measured.  A spectroscopic survey of the field has revealed several
tens of galaxies at the redshift of the lens (McKean et al., in prep).

\section{X-ray Data Reduction and Analysis}

\subsection{X-ray Data Reduction}

\label{xray-data}

In this section, we present results obtained from {\em Chandra} observations
of the groups described in the previous section.  We report the first
results for the groups associated with B0712 and B2108, while
the data associated with the other groups have been obtained from the 
{\em Chandra} archive and reprocessed so that the full sample has been
processed in an identical manner.  We will compare the results from
the reprocessed data with those obtained from earlier work 
by \citet[][for the PG~1115 and B1422 systems]{grant_chandra}
and \citet[][for the B1600 and B1608 systems]{chandra1608}.

Observations of B0712 and B2108 were carried out with {\it Chandra's}
Advanced CCD Imaging Spectrometer \citep[ACIS;][]{acis} on 2003
December 17 (ObsID 4199) and 2006 July 14 (ObsID 6971), respectively.
Both fields were imaged with the nominal 3.2 s CCD frame time for a
total integration of 45.5 ks for B2108 and 97.7 ks for B0712.  The
resulting data were transmitted in VFAINT mode for both observations.
An examination of light curves produced from the datasets in the
0.3-10 keV band shows no indication of flaring during either
observation.  Both targets were imaged with the back-illuminated
ACIS-S3 chip, with B0712 and B2108 located $44^{\prime\prime}$ and
$75^{\prime\prime}$, respectively, from the aim point of the
observation.

Data for the PG1115, B1422, B1600, and B1608 systems were obtained
from the {\it Chandra} archive maintained by the {\it Chandra} X-ray
Center (CXC)\footnote{http://cxc.harvard.edu/}.  The observational
parameters for these datasets are listed in Table
\ref{obs-param-table}, along with those of the B0712 and B2108
observations.  All of the archival fields were imaged with the ACIS-S3 chip
near the telescope aimpoint, with the largest off-axis observation
being that of B1608, imaged roughly 75$^{\prime\prime}$ from the
aimpoint.  The observations of PG1115 and B1422 were carried out in FAINT
mode, while those of B1600 and B1608 are in VFAINT mode; all four
employed the standard CCD frame time of 3.2 s.  We searched for
flaring events in the light curves of each observation using the {\tt
lc\_clean.sl} script and detected periods of increased background in
only the 367 dataset of B1422.  Excluding periods in which the
background count rate differed from the mean rate by a factor greater
than 1.5 reduced the usable exposure time of that observation from
28429 to 16888 seconds.

\ifsubmode 
\clearpage
\begin{figure}[t]
\epsscale{.8}
\else 
\begin{figure*}[t] 
\fi 
\plotone{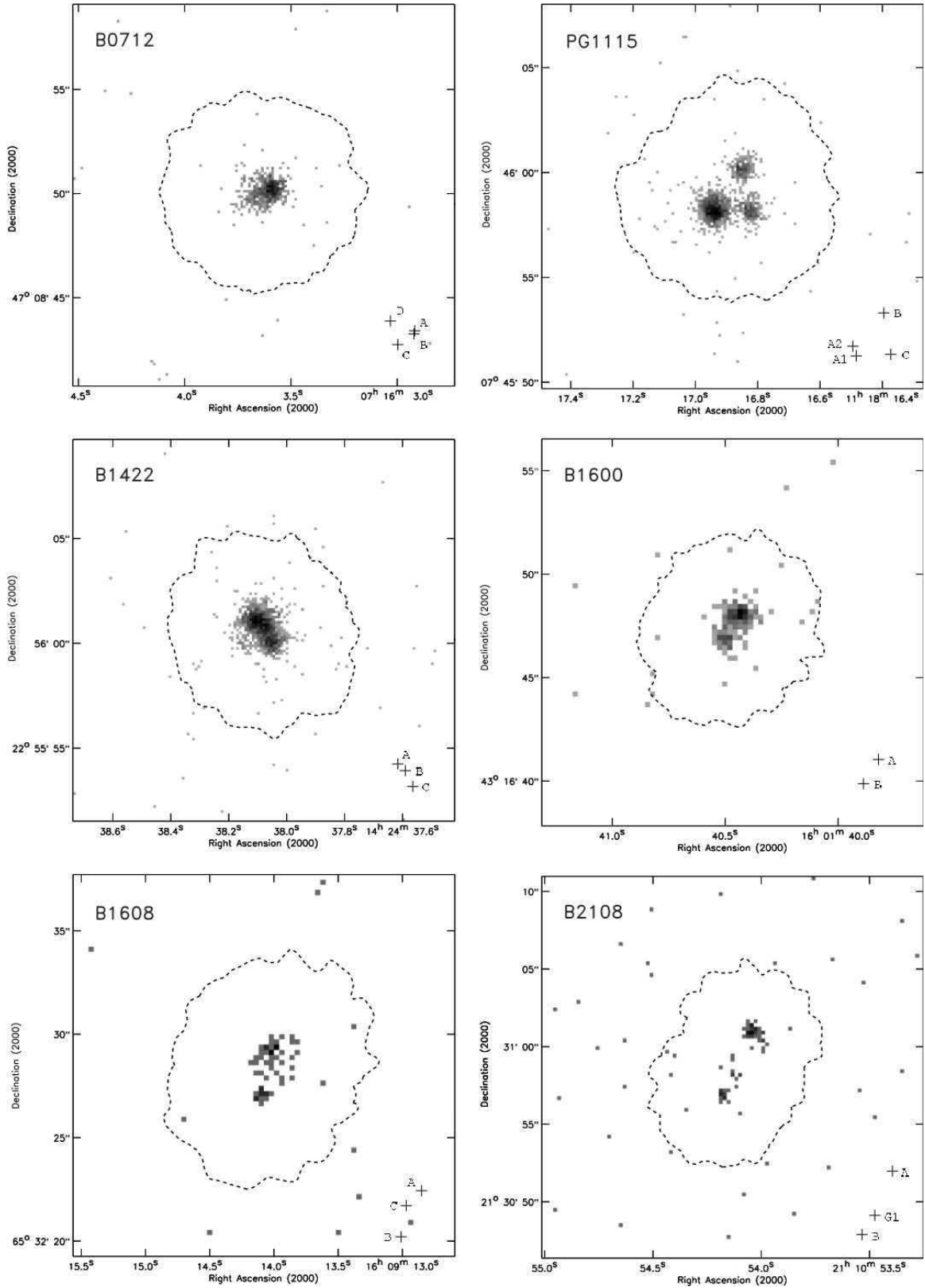}
\caption{
Soft band (0.5-2.0 keV) X-ray images of the six lens systems, binned
to a resolution of 0$\farcs$123.  The dashed line denotes the masking
aperture constructed to contain $99\%$ of the flux from each set of
point sources.  
The schematic in the lower right corner of each panel
represents the configuration of the lens system at radio
and optical wavelengths.  For each case, a correspondence between
the X-ray and optical/radio morphologies can be seen.
%
}
\label{fig_rawximg}
\ifsubmode 
\end{figure}
\clearpage
\else 
\end{figure*} 
\fi 

All six datasets were reprocessed and analyzed using standard Chandra
Interactive Analysis of Observations (CIAO) 3.3 software tools and
version 3.2.2 of the Chandra calibration database available through
CXC.  We produced new bad pixel masks for the 363, 367, and 4199
datasets with the {\tt acis\_run\_hotpix} script, as the original
masks were created by an older version of the CXC pipeline which
misidentified afterglow events and failed to detect hot pixels.  New
level 1 event files were produced for all of the observations using
the {\tt acis\_process\_events} script, which makes use of the latest
gain files and corrects for the effects of time-dependent gain
variations and charge transfer inefficiencies (CTI) in the ACIS CCDs.
To improve image quality the preprocessing was implemented without
event pixel randomization.  Level 2 event files were produced by
filtering on standard ASCA grades (grades=0,2,3,4,6), good status bits
(status=0), and Good Time Intervals (GTIs) supplied by the CXC
pipeline.  To examine the extended emission originating from the
galaxy groups associated with each system we produced images in the
soft X-ray band (0.5-2 keV), where emission from the intragroup gas
would have the greatest signal.  Each image was corrected for
vignetting using exposure maps created at an energy of 1.5 keV.  The
pixel-specific vignette correction factor is estimated by normalizing
the exposure map to its maximum value at the aimpoint of the
observation.  Images of the lens systems, binned to a pixel scale of
0$\farcs$12, are shown in Figure \ref{fig_rawximg}.  We note that the
sub-pixel binning is used only to make Figure~\ref{fig_rawximg}; all
analysis is done with standard-size pixels.

\ifsubmode 
\clearpage
\fi 
\begin{deluxetable}{lrlll}
\tablewidth{0pt}
\tablecaption{Observation Parameters}
\tablecolumns{5}
\tablehead{  
\colhead{Target}
 & \colhead{Obs ID}
 & \colhead{Exp (s)}
 & \colhead{Obs Date}
 & \colhead{PI}}
\startdata
B0712 &  4199  &  97742  &  2004 Dec 17  &  Fassnacht \\
PG1115 & 363   &  26489  &  2000 Jun 2   &  Garmire   \\
B1422  & 367   &  16888\tablenotemark{a}  &  2000 Jun 2   &  Garmire    \\
B1600  & 3460  &  30176  &  2003 Oct 7   &  Kochanek  \\  
B1608  & 3461  &  29717  &  2003 Sep 9   &  Kochanek  \\
B2108 &  6971  &  45507  &  2006 Jul 14  &  Fassnacht \\
\enddata
\tablenotetext{a}{Reduced from 28429 seconds due to flaring events.}
\label{obs-param-table}
\end{deluxetable}
\ifsubmode 
\clearpage
\fi 

\subsection{X-ray Analysis}

In order to characterize the X-ray properties of the group sample, we
must first remove any point-like emission in the field of each system.
This includes emission from the lensed active galactic nuclei (AGN),
and possibly from the lensing galaxies, which are expected to be
embedded in the fainter diffuse group emission.  The excellent spatial
resolution of {\em Chandra} facilitates the crucial separation of the
point-source images from the underlying diffuse emission.  Raw,
soft-band images of the lens systems are shown in
Figure~\ref{fig_rawximg}.  
For each system, visual inspection reveals morphologies similar to those
seen at other wavelengths, allowing the straightforward registration of 
the X-ray frame to the astrometry of the existing radio and optical
data.
%
In the B0712 system, the observed X-ray emission corresponds to the
three brightest images of this quadruply lensed source,
although the individual images are largely blended because their
maximum separation is only 1$\farcs$5.  The three components A, B, and
C in PG1115 are well separated but the two images A1 and A2 have not
been resolved.  In B1422 we clearly see the blended emission from
images A, B, and C, but not the fainter image D.  The B1600 morphology
shows images A and B, while in B1608 components A and C are blended
and only image B is well separated; image D is not detected.  The
images of B1600 and B1608 match well the images presented by
\citet{chandra1608}, as expected.  In B2108 we see emission from both
lensed images and the lensing galaxy (G1). The lensing galaxy in this
system is therefore loud in both X-rays and radio.

\ifsubmode 
\clearpage
\begin{figure}[t]
\epsscale{0.8}
\else 
\begin{figure*}[t] 
\fi 
\plotone{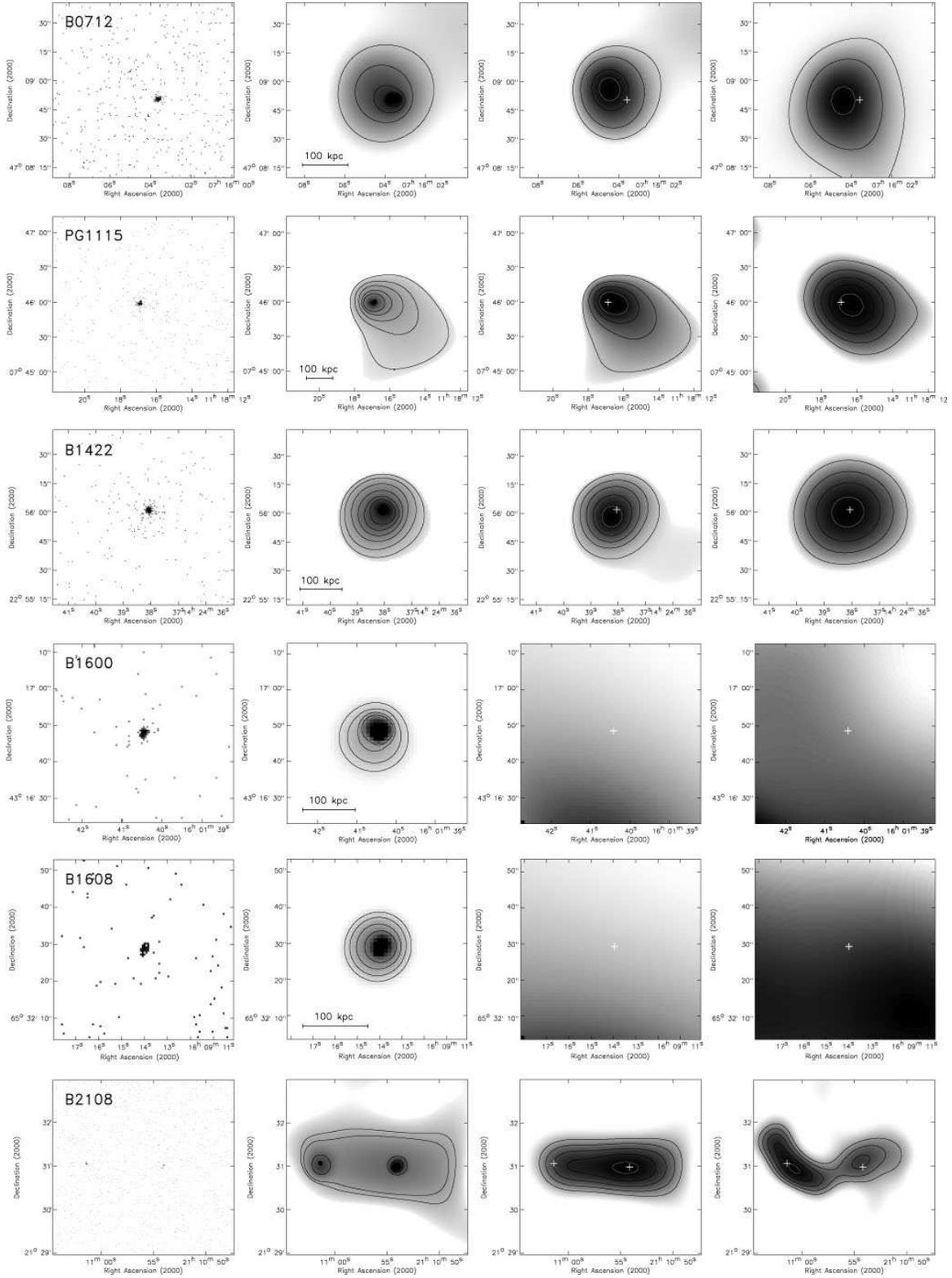}
\caption{
{\em Column 1:} Raw soft band (0.5-2.0 keV) X-ray images binned to a pixel
scale of 0$\farcs$492 of the field of each galaxy group.
{\em Column 2:} Adaptively smoothed images of each system prior to 
the masking process.  
{\em Column 3:} Adaptively smoothed images of each system following
the masking process.  The position of the masked
point sources are marked with plus signs.  The flaring seen on the
right edge of the B2108 adaptively-smoothed image is due to the
chip edge.
{\em Column 4:} Same as column 3, but smoothed with a fixed-width 
20\arcsec\ Gaussian kernel.
The contour levels in the smoothed images are chosen to highlight the
important structures.  The scale bars in the second column are
drawn under the assumption that $h = 0.7$.
}
\label{fig_smoothximg}
\ifsubmode 
\end{figure}
\else 
\end{figure*} 
\fi 

Removing the contribution of these lensed images is complicated as
there is significant structure and signal in the wings of the {\em
Chandra} point spread function (PSF) even at low off-axis angles.
Care must be taken to prevent residual flux in the wings from
artificially enhancing any group component.  While previous studies
have used complex two-dimensional models to disentangle the point-like
and diffuse emission \citep{grant_chandra,pooley06} we have employed a
relatively simple masking technique that uses PSF modeling to quantify
the extent of point sources in each field and replace the point
sources with an estimate of the local background.  In order to
minimize any contamination from the wings of the PSF, our masking
apertures were constructed to contain 99\% of the flux from a given
point source.  Using a process similar to that employed by the {\tt
ACIS Extract} \citep{acis_extract} package, model PSFs were
constructed using the CIAO tool {\tt mkpsf} at an energy of 1.5 keV.
As the Chandra PSF is dependent on off-axis angle, a unique PSF was
constructed for each lensed image.
Once the X-ray data had been
registered to the optical and radio frames, the positions of the
lensed components were taken from the literature and used as the
centroids for the masking apertures.  
%
We draw from the 5 GHz and 8.4
GHz radio observations of \citet{0712disc}, \citet{patnaik99},
\citet{1608delays2}, and \citet{2108disc} for B0712, B1422,
B1608, and B2108, respectively, and from the {\it Hubble Space
Telescope} (HST) optical observations of \citet{impey98} for PG1115.
Each model PSF is then convolved with a two-dimensional Gaussian with
$\sigma=0\farcs27$ to account for the telescope dither blur
\citep{acis_extract} and a 99\% enclosed energy contour is constructed
from the smoothed image.  The outer extent of the combined contours
for a set of lensed images is then defined as our masking aperture for
that system.  These masking regions are shown plotted on their
respective fields in Figure
\ref{fig_rawximg}.  A local background is then determined directly outside
the masking aperture and the masked region is filled with the median
background level, scaled to the area of the masked region, using the {\tt
dmfilth} task.

\ifsubmode 
\clearpage
\begin{figure}[t]
\else 
\begin{figure*}[t] 
\fi 
\epsscale{1.0}
\plotone{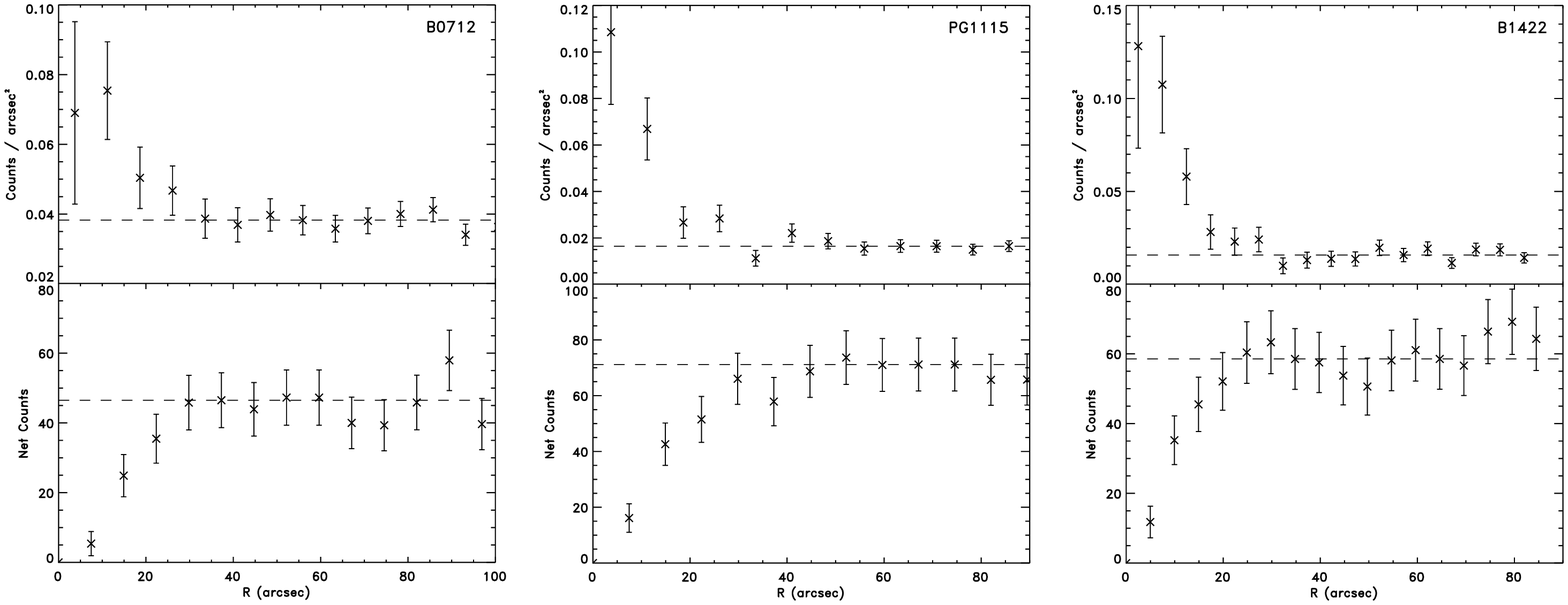}
\caption{
\emph{Upper Panels:} 
Azimuthally averaged surface brightness profiles for the B0712,
PG1115, and B1422 groups, constructed following the masking of point
sources in the field.  The background level of each field is denoted
by the horizontal dashed line.  
\emph{Lower Panels:} 
Cumulative net count profiles for each system detected above the
background.  The total number of counts originating from each group is
shown by the horizontal dashed line.  
}
\label{gcafig1}
\ifsubmode 
\end{figure}
\clearpage
\else 
\end{figure*} 
\fi 

The images were next smoothed in order to emphasize
the diffuse X-ray emission associated with these systems.
Both adaptive smoothing, using the CIAO tool {\tt csmooth}
\citep{asmooth}, and fixed-width Gaussian smoothing techniques
were used.   The results are shown in Figure \ref{fig_smoothximg}. 
The middle two columns of the figure show the
%
adaptively-smoothed images of each system prior to and following the
masking process. In many of the fields a diffuse component is clearly
visible even with the lensed images present.  After masking these
sources, we applied the adaptive smoothing algorithm with minimum and
maximum significance thresholds of 3$\sigma$ and 5$\sigma$,
respectively.  The minimum and maximum smoothing scales were allowed
to float.  An examination of the resulting images showed diffuse
emission in four of the six fields: B0712, PG1115, B1422, and B2108.
On the other hand, we see no obvious signal in the fields of
B1600 and B1608, in agreement with the results of \citet{chandra1608}
who also fail to detect any significant extended emission within
4\arcmin\ of the lensed galaxies.
%
In the B1608 field, we do notice a region of enhanced emission near to
the location of the B1608-3 group.  However, this location is close to
the chip edge, and it is not clear if the emission is real or due to
a higher background level.  Deeper X-ray observations are needed to
assess whether the B1608-3 group has associated diffuse X-ray emission.

%
The right-most column of Figure \ref{fig_smoothximg} shows the
masked data smoothed by a fixed-width Gaussian with a kernel size of
20\arcsec.  For all but one of the lens systems, the fixed-width
smoothing produces an image that strongly resembles the image produced
from the adaptive smoothing.  This is clearly not the case for the
B2108 field, where the fixed-width smoothing reveals at least two
major components.  The difference between the adaptive and fixed-width
smoothing results may be due to the bridge of relatively bright
emission connecting the two components.  We believe that the adaptive
smoothing algorithm probably detected the two components plus the
bridge as a single large object and thus incorrectly smoothed the data
with a very large kernel size.  To assist in the interpretation of the
smoothed images, we binned the raw masked data to 16\arcsec\ pixels.
The binned imaged shows a somewhat U-shaped structure that more
closely resembles the Gaussian-smoothed image than the
adaptively-smoothed image.  We therefore feel that for B2108 the
fixed-width smoothing has probably produced a more accurate
representation of reality.  We will, thus, use the Gaussian-smoothed
image in the following discussion.  We designate the western
component, which is the one roughly centered on the lens system, as
B2108-1.  The eastern component will be referred to as B2108-2.
%

The majority of the quantitative results, such as the total
counts, fluxes, etc., that are presented below are
derived from the masked unsmoothed images.  However, the smoothed
images are used to determine the centroids of the diffuse X-ray
emission.  Because we feel that the fixed-width Gaussian smoothing
produces a more realistic representation of the B2108 emission, and
because the fixed-width smoothing also seems reasonable for the other
lens fields, we use the Gaussian smoothing to determine the centroids
for all of the fields.  The resulting centroids are listed in Table 2.
%
It is worth
noting that the lack of structure in the smoothed images of B1600 and
B1608 suggests that there is no significant residual flux or artifacts
produced as a result of the masking process itself; we proceed under
the assumption that this holds true for the other four fields as well.

\begin{figure}[t]
\epsscale{0.7}
\plotone{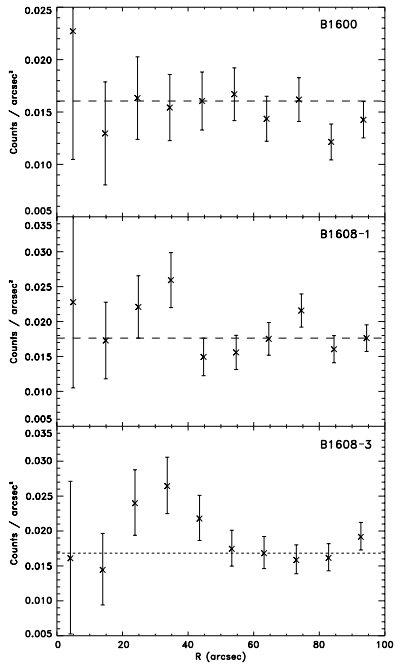}
\caption{Azimuthally averaged surface brightness profiles for the
B1600 and B1608 fields.  Because no significant diffuse emission is
detected at these locations, the profiles are centered on the optical
centroids of the groups.}
\label{gcafig2}
\end{figure}

It was not possible to do a full spectroscopic analysis for any
of the fields, due to the low number of counts produced by the diffuse
emission.  Although an analysis of the B2108-1 system, which has the
highest number of counts among the detected systems, did yield a
temperature, the uncertainties were so large as to render the value
meaningless.  Therefore, 
%
we instead determined the soft-band flux of the group emission by
normalizing a Raymond-Smith spectral model in the CIAO package {\tt
Sherpa} to the net counts detected above the background in each
system.  The net counts were measured using a standard growth-curve
analysis, 
%
although the analysis of B2108 was more complex than that of the
other systems.  For all but the B2108 system, the local background
levels were determined by creating azimuthally-averaged surface
brightness profiles.  For the systems with detected diffuse emission,
the annuli were centered on the peak of the emission.  For the B1600
and B1608 systems the annuli were placed at the group centers as
determined from optical observations.  The background for each system
was set to the median surface brightness in the region where the
profile leveled off.  
%
The widths of the annuli were held constant for
a given system, but varied between 5\arcsec\ and 10\arcsec\ from field
to field, depending on the achieved signal-to-noise ratio of the
detected diffuse emission.  
%
The surface brightness profiles are shown in Figures~\ref{gcafig1}
and \ref{gcafig2}.
In the case of the two components seen in the B2108 field, the standard
approach could not be used because the complex morphology of the emission
produces cross contamination of the group counts.  Therefore, instead
of using surface brightness profiles we set the background level by
measuring the surface brightness in a source-free area of the chip.
%

\ifsubmode 
\clearpage
\fi 

\begin{figure}[t]
\epsscale{1.0}
\plotone{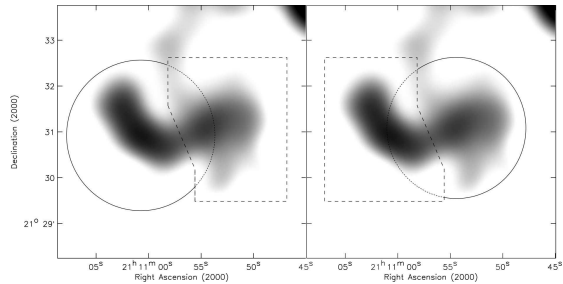}
\caption{
Gaussian-smoothed images of the B2108 field showing the masking
regions (dashed polygons) used to minimize the cross-contamination of
the counts associated with each component.  Counts in these masking regions
were excluded when measuring the cumulative net profiles of the
components.  The circle in each panel is centered on the component
for which the counts are being measured, and shows the radius at which the
background level was reached.
}
\label{fig_b2108masks}
\end{figure}

\begin{figure}[t]
\epsscale{0.8}
\plotone{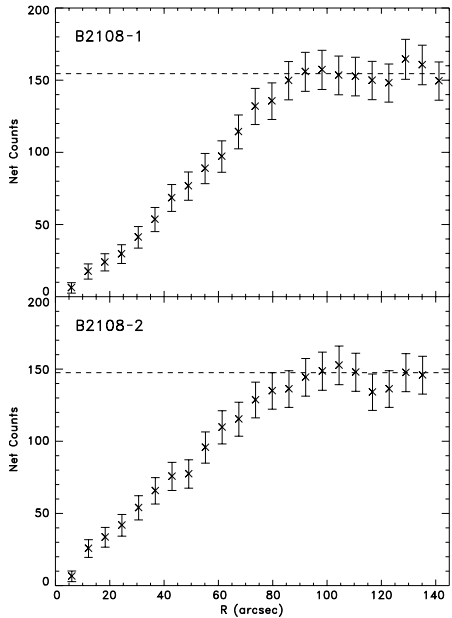}
\caption{Cumulative net count profiles for the two regions of diffuse
emission in the B2108 field.  The masking that was applied to reduce
cross contamination between the groups will lead to underestimated
count levels in the outer regions of the profiles.}
\label{gcafig3}
\end{figure}

A cumulative net count profile for each field was constructed by measuring the
counts in successively larger apertures centered on the group emission
and subtracting an appropriately scaled background.  We take the total
number of counts originating from the group to be the level at which
the cumulative profile ceases to grow.  
%
In the case of the B2108 components, it was necessary to mask out
parts of the apertures where the counts from the group in question
were significantly contaminated by the emission from the other group.
Figure~\ref{fig_b2108masks} shows the masking regions used and 
Figure~\ref{gcafig3} shows the resulting curves of growth.
%
The total counts measured in each system are listed in Table 2, 
%
where the counts from the B2108 will be underestimates due to the masking
procedure.
%
We find significant signal above the background for the B0712, PG1115,
B1422, and B2108 systems, with the emission in the B0712 detected at
the $3.5~\sigma$ level and the other systems all being detected at the
$4.5-5~\sigma$ level.  We did not detect significant emission in the
B1600 and B1608 systems, confirming our non-detection in these fields
from the image analysis.  As discussed below, the fluxes we present
for the B1600 and B1608 systems are $3\sigma$ upper limits.

 \ifsubmode 
\clearpage
\fi 
\ifsubmode 
\clearpage
\fi 

For the four fields with detected diffuse emission, we converted from
net counts to flux by modeling the group emission in {\tt Sherpa} as a
Raymond-Smith thermal plasma with a metal abundance of $0.3 Z_{\odot}$
and a gas temperature of 1 keV.  The instrument response is taken into
account by creating redistribution-matrix and auxiliary-response files
with the {\tt specextract} task for each observation.  Using these
response matrices and the Raymond-Smith model, along with the redshift
for the groups and the exposure times of the observations, we
determined the source flux required to produce the observed counts in
each field.  These fluxes were then corrected for the effects of
Galactic absorption using the neutral hydrogen column densities of
\citet{dickey-lockman90}.  Finally, these fluxes were converted to
rest-frame soft-band luminosities and from there to rest-frame
bolometric luminosities.  The computed fluxes and luminosities are
given as columns 7 and 8 in Table~\ref{tab_xray}.

For ease of comparison with other determinations of group X-ray
properties, we also computed values within apertures of radius
$R_{500}$.  
The value of $R_{500}$ for each system was estimated
from the radial velocity dispersion that had been determined from the
group galaxy redshifts (Table 3), as 
$R_{500} = 2 \sigma_v /[ \sqrt{500} H(z)]$.
%
For each group, we had to extrapolate from the region of observed
emission out to $R_{500}$.  To do this, we used a $\beta$ model with
$\beta = 2/3$ and $R_{\rm core} = 160$~kpc; these are the median and
mean values, respectively, from the fits to the intermediate-redshift
group sample of \citet{jeltema07}.  The rest-frame bolometric
luminosity within $R_{500}$ is given as the last column in
Table~\ref{tab_xray}.
To obtain upper limits on the flux in the
fields of B1600 and B1608 we used a similar process but instead
normalized the spectral models to the background counts within
apertures of radius $R_{500}$ centered on the lensed images.  The
values listed in Table~\ref{tab_xray} for these three groups are the
$3\sigma$ upper limits within these apertures.

\ifsubmode 
\clearpage
\fi 

\section{Optical Data Reduction and Analysis}

In order to compare the X-ray and optical properties of the group
sample, we require estimates of the group velocity dispersions.  As we
did for the X-ray properties, we calculate the velocity dispersions
for all the groups using a standardized approach in order to minimize
effects due to different computation methods.  To do this we have
taken the updated redshift distributions for the PG1115 and B1422
systems from \citet{momcheva} and the B1600 redshift information from
\citet{auger_groups1}.  Furthermore, we have supplemented the
previously existing redshift data on B0712 from \citet{0712group} and
on B1608 from \citet{1608groupdisc} with new data that we
present below.  Finally, we include the preliminary analysis of the
data for the B2108 field.  These data will be presented fully by
McKean et al.\ (in prep).

\subsection{B0712 Spectroscopy}

\begin{figure}
\epsscale{1.0}
\plotone{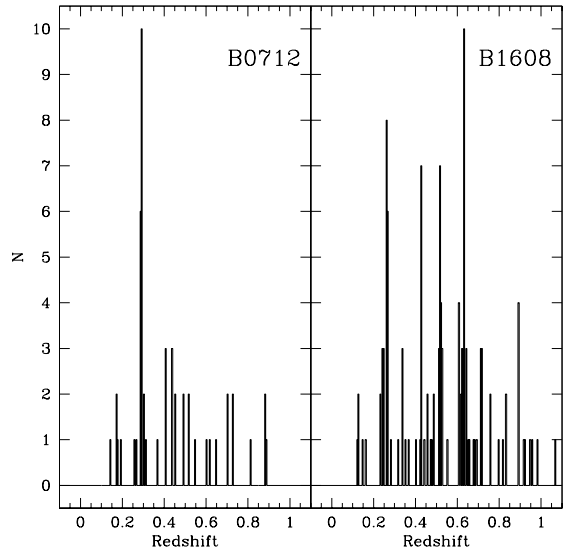}
\caption{
Galaxy redshift distributions in the B0712 ({\em left}) and B1608 ({\em right})
fields.  The bins have width $\Delta z = 0.005$.
}
\label{fig_zhist}
\end{figure}

Previous observations of the environment of the B0712 field yielded
34 non-stellar redshifts, with 10 galaxies comprising the foreground
group \citep{0712group}.  We obtained further spectroscopy on the field
on 2004 April 10 with the Low Resolution Imaging Spectrograph
\citep[LRIS;][]{lris} on the Keck I Telescope.  The observations
consisted of three multislit exposures of 1800~sec each.  Arc lamp and
internal flat-field exposures were obtained following the science
exposures.  The data were obtained with both the red and blue LRIS
cameras, with the D560 dichroic splitting the incoming beam at
$\sim$5700\AA.  The red-side data were dispersed by the 600/7500 grating,
giving a nominal scale of 1.28~\AA~pix$^{-1}$.  On the blue side
the 400/3400 grism was used, providing a nominal 1.09~\AA~pix$^{-1}$\ 
dispersion.  The data were reduced with an python-based multislit
reduction package developed by M.\ Auger.  This package detects the
slits, does the bias subtraction and flat-field correction, corrects
for distortions in the spatial direction, does the wavelength
solution, rectifies the spectra, subtracts the sky, detects objects in
each slit, and extracts the spectra, all automatically.  The slitmask
had 32 slits, from which 13 non-stellar spectra were extracted.  
The updated redshift distribution is shown in Figure~\ref{fig_zhist}a.

\ifsubmode 
\clearpage
\fi 
\ifsubmode 
\clearpage
\fi 

\subsection{B1608 Spectroscopy}

The spectroscopic data on the B1608+656 field that are presented in 
\citet{1608groupdisc} yielded 97 non-stellar redshifts, in a distribution
that showed four clear spikes.  Additional multislit observations of
the field were obtained on 2004 Aug 13 and 2007 June 12, with one
slitmask being used on each occasion.  In both cases, both red- and
blue-side data were obtained.  The first set of observations used the
D560 dichroic, the 600/7500 red-side grating, and the 600/4000
blue-side grism, yielding nominal dispersions of 1.28 and 0.63
\AA\ pix$^{-1}$ for the red and blue sides, respectively.  Two 
1800~sec observations were obtained.  The second set used the D680
dichroic, the 831/8200 grating, and the 300/5000 grism.  This
combination provided nominal dispersions of 0.93 and 1.43
\AA\ pix$^{-1}$.  This second set of observations was designed to
measure an improved stellar velocity dispersion for the B1608
primary lensing galaxy and to make the first measurements of the
stellar velocity dispersions of the two additional strong lens
candidates in this field \citep{1608serendip}, so the total exposure
times were long.  In total, 11 exposures of length 1800~s were
obtained through this mask.  Both masks were reduced using the
automated python pipeline. The observing conditions during the 2004
observations were substandard and only two new redshifts were measured
from the mask.  In contrast, 26 new non-stellar redshifts were
measured from the 2007 observations.  The updated redshift
distribution is shown in Figure~\ref{fig_zhist}b.

\ifsubmode 
\clearpage
\begin{deluxetable}{lcrrrrlccc}
\rotate
\else 
\begin{deluxetable*}{lcrrrrlccc} 
\fi 
\tabletypesize{\scriptsize}
\tablewidth{0pt}
\tablecaption{X-ray Properties of Diffuse Group Emission}
\tablecolumns{9}
\tablehead{  
 \colhead{} 
 & \colhead{RA}    
 & \colhead{Dec}   
 & \colhead{$N_{\rm H}$}
 & \colhead{Net} 
 & \colhead{Count Rate} 
 & \colhead{Flux\tablenotemark{a,b}}
 & \colhead{$L_{\rm X,bol}$\tablenotemark{b}}
 & \colhead{$L_{\rm X,bol,500}$\tablenotemark{b}}
 \\
 \colhead{Group}
 & \colhead{(J2000)}
 & \colhead{(J2000)}
 & \colhead{($10^{20}$ cm$^{-2}$)}
 & \colhead{Counts} 
 & \colhead{($10^{-3}$ s$^{-1}$)} 
 & \colhead{($10^{-15}$ erg cm$^{-2}$ s$^{-1}$)}
 & \colhead{($10^{42} h^{-2}$ erg s$^{-1}$)}
 & \colhead{($10^{42} h^{-2}$ erg s$^{-1}$)}
}
\startdata
B0712    & 07:16:04.4  & +47:08:49  &  7.8 &       47 & 0.48 
 & $  1.6 \pm 0.2 $  & $  0.44 \pm 0.05 $ & $  1.3 \pm 0.2 $ \\
PG1115   & 11:18:16.3  & +07:45:57  &  4.0 &       71 & 2.7  
 & $  6.9 \pm 0.8 $  & $  2.2 \pm 0.3 $ & $  3.9 \pm 0.5 $ \\
B1422    & 14:24:38.1  & +22:56:00  &  2.7 &       59 & 3.5  
 & $  9 \pm 1 $      & $  3.4 \pm 0.5 $ & $ 12  \pm 1 $ \\
B1600    &    \nodata  &    \nodata &  1.3 &  \nodata & \nodata 
 & $ <1.8$           & $ <1.2 $  & $ <1.2 $   \\  
B1608-1  &    \nodata  &    \nodata &  2.7 &  \nodata & \nodata 
 & $ <2.0$           & $ <4.0 $ & $ <4.0 $ \\
B1608-3  &    \nodata  &    \nodata &  2.7 &  \nodata & \nodata 
 & $ <4.3$           & $ <3.0 $   & $ <3.0 $   \\ 
B2108-1  & 21:10:54.4  & +21:31:05  & 12.6 &      155 & 3.4  
 & $ 12 \pm 1 $      & $ 5.9 \pm 0.5 $ &  $5.9 \pm 0.5 $ \\
B2108-2  & 21:11:00.7  & +21:30:55  & 12.6 &      148 & 3.3  
 & $ 11 \pm 1 $      & $ 5.4 \pm 0.5 $ &  \nodata \\
\enddata
\label{tab_xray}
\tablenotetext{a}{0.5-2.0 keV}
\tablenotetext{b}{Corrected for Galactic Absorption}
\ifsubmode 
\end{deluxetable}
\clearpage
\else 
\end{deluxetable*} 
\fi 

\subsection{Optical Analysis}

In order to minimize the likelihood of spurious conclusions arising
from different data analysis methods, we computed the group velocity
dispersions for each of the systems in the same way.  In some cases
(B1422, PG1115), this means re-analyzing the data from the literature,
while for others (B0712, B1608, B2108) we have acquired new data which
have been combined with previously published data, if available.  For
the B1600 group, the published group parameters
\citep{auger_groups1} were computed using our standardized technique
and therefore they were taken directly from the literature.  The first
step is to identify the groups from the redshift distributions.  To do
this, we follow the iterative group-finding procedure presented in
\citet{auger_groups1}.  In some cases, this objective method leads to
slightly different membership than that presented in the literature.
The number of galaxies given for each of the groups in
Table~\ref{tab_optdata} reflects the numbers from the current
analysis.  We note that for the B1608 field, the objective group
finder identifies the same four redshift spikes as noted in
\citet{1608groupdisc}, plus an additional group candidate at
$z = 0.71$.  However, we only consider two groups in our analysis: the
one that is physically associated with the lensing galaxy (B1608-1) and
the $z = 0.426$ group (B1608-3).  Both of these groups are compact
spatially and in redshift, and are clearly centered in the
region covered by our data.  In contrast, the $z = 0.26$ group appears
to be real, but also has a spatial distribution that suggests that it
may be centered  off the region covered by the spectroscopic and X-ray
data.  The $z = 0.52$ spike has a filamentary spatial distribution and
is composed mostly of late-type galaxies, so we do not believe that it
is a real group.  The new $z = 0.71$ candidate has too few members for
us to accurately characterize its properties at this time, and is also
so distant that the relatively shallow {\em Chandra} data do not
provide interesting constraints on its properties.

The velocity dispersions were calculated from the distributions of the
redshifts of the identified group members, using the methods described
in \citet{beers90}.  The gapper algorithm was used for groups with
fewer than 15 members, while for groups with more members we used the
bi-weight estimator.  In each case, the errors on the resulting
velocity dispersions were determined using a bootstrap approach.  The
dispersions and their errors are given in Table~\ref{tab_optdata}.
Although the group velocity dispersions have been calculated in a
standard manner, we note that a dispersion determined from only
$\sim$10 members may be strongly biased with respect to the true value
\citep[e.g.,][]{zm98,gal07} and thus the calculated values
should be used with care.

\ifsubmode 
\clearpage
\fi 
\begin{deluxetable}{lrrrrr}
\tabletypesize{\scriptsize}
\tablecolumns{10}
\tablewidth{0pc}
\tablecaption{Group optical properties}
\tablehead{
\colhead{}
 & \colhead{}
 & \colhead{Group}
 & \colhead{$\sigma_v$}
 & \colhead{$R_{500}$}
 & \colhead{$\theta_{500}$}
\\
\colhead{Group}
 & \colhead{$N$}
 & \colhead{Redshift}
 & \colhead{(\kms)}
 & \colhead{($h^{-1}$\ kpc)}
 & \colhead{(arcsec)}
}
\startdata
B0712   & 15 & 0.290 & 320$\pm$20 & 250 &  82 \\
PG1115  & 13 & 0.310 & 450$\pm$70 & 340 & 110 \\
B1422   & 16 & 0.339 & 460$\pm$90 & 350 & 105 \\
B1600   &  7 & 0.415 &  90$\pm$20 &  72 &  19 \\
B1608-1 & 10 & 0.632 & 150$\pm$30 &  95 &  20 \\
B1608-3 &  8 & 0.426 & 320$\pm$90 & 190 &  49 \\
B2108   & 47 & 0.364 & 470$\pm$50 & 350 &  98 \\
\enddata
\label{tab_optdata}
\end{deluxetable}
\ifsubmode 
\clearpage
\fi 

\section{Discussion}

\subsection{Comparison to Previous X-ray Analyses}

The diffuse X-ray emission from two of the lens-selected groups,
PG1115 and B1422, has been analyzed previously by
\citet{grant_chandra}.  Because our analysis uses different inputs and
techniques, care should be taken when comparing the results of the two
analyses.  As one example, each of these two systems was observed
under two separate programs (Obs IDs), namely 363 and 1630 for PG1115,
and 367 and 1631 for B1422.  To simplify the analysis of the effects
of the PSF, we used only the longer of the two
observations for each system.  In contrast, the \citet{grant_chandra}
analysis combined the two programs in each case.  Therefore, the
germane basic quantity to use in the comparison of the results is not
the net counts from the diffuse emission, but rather the net count
rates.  In each case, we find a higher count rate than that measured
in the previous analysis: $2.7 \times 10^{-3}$ vs.\ $1.3 \times
10^{-3}$ for PG1115 and $3.5 \times 10^{-3}$ vs.\ $1.8 \times 10^{-3}$
for B1422, where all values are in counts s$^{-1}$.  We believe that
the cause of this discrepancy is the manner in which we masked out the
lensed AGN emission for the lenses.  The
\citet{grant_chandra} approach was conservative, masking out regions
with diameters of 14\arcsec--16\arcsec, while our 99\% masking regions
(Fig.~\ref{fig_rawximg}) cover somewhat smaller areas, with typical
sizes of $\sim$10\arcsec\ across.  In both cases, a correction was
then made to account for the diffuse flux within the masked region.
Given the small angular extent of the groups, changes in the masking area
may lead to differences in the size of the correction and therefore to
significant changes in the measured flux of the diffuse emission.

A second difference in technique is that 
the luminosities that
we use in our final analysis are those within $R_{500}$, rather than
just using the luminosities calculated from the observed region of
significant detection.
To facilitate comparison with the \citet{grant_chandra} results, the
bolometric rest-frame luminosities in Table~\ref{tab_xray} are given
for both the detection region and the $R_{500}$ region.  Other
differences arise from the X-ray temperature used for PG1115, where
\citet{grant_chandra} use 0.7~keV and we use 1.0~keV, and the assumed
cosmology.  The \citet{grant_chandra} results are computed using an
Einstein-de Sitter cosmology with $h = 0.5$, while we use
$(\Omega_{\rm M},\Omega_\Lambda) = (0.3,0.7)$ and set $h = 0.7$ when
we have to fix its value.

It is important to consider the effects that the analysis techniques
will have on the interpretation of the results.  In terms of many of
the conclusions drawn from the measured X-ray luminosities, including
the discussion of the $L_X$--$\sigma_v$ relationship below, the
factors of a few in $L_X$ that come out of the different techniques
will not change the interpretation.  In particular, given the log-log
nature of the $L_X$--$\sigma_v$ plot and the scatter in the observed
relationship, these factors of 2--3 do not significantly change the
location of the lens-group points in Figure~\ref{fig_lx_sig}.  On the
other hand, the determination of the location of the X-ray centroid
is more sensitive to how the lensed AGN emission is masked, and
therefore conclusions about the possible offsets of the brightest
group galaxies from the center of the X-ray emission are tentative at
best.

\subsection{The Nature of the X-ray Emission in B2108}

The X-ray emission associated with the B2108 system is shown in
the last row of Figure~\ref{fig_smoothximg}, where its morphology is
seen to be clearly different from that of the other systems.  While
the smoothed data in each of the other fields show a single,
relatively compact region, the B2108 field contains two elongated
regions, connected by a bridge of emission.  These regions are roughly
centered on the positions of masked point sources; the western
component (B2108-1) is the one centered on the lens system.  

It is likely that the X-ray emission in B2108-1 is produced by
the the group associated with the B2108+213 lens system.  The lensing
galaxy is a massive elliptical, is the brightest galaxy in the group,
and is found nearly at the center of the X-ray emission.  However, 
the source of emission in B2108-2, the eastern X-ray component, is
much less clear.  
An examination of optical imaging of this field, obtained with the Keck
telescope, shows a $R>23$ source coincident with the X-ray position of
the 
%
point source that is roughly at the center of B2108-2.  
%
The faintness of this optical object compared to the confirmed group
members ($\sim 5$ magnitudes fainter than the primary lensing galaxy)
suggests that it is background AGN rather than a source associated with a
massive elliptical in the $z = 0.36$ group.  If this is the case, 
%
then it is not clear whether the diffuse X-ray emission in
B2108-2 is coming from a massive system associated with a background
AGN or whether it is instead due to a second group at the redshift
of the lensing galaxy.  
%
The clear removal of the point-source emission
for the B1600 and B1608 systems suggests that the
%
diffuse emission
%
is not wholly due to residual emission from the point source.  
%
The majority of the B2108-2 emission is outside the region for
which spectroscopic data have been obtained, so there is no
information on whether there may be an overdensity of higher-redshift
objects in the area covered by this second system.  That being said,
the available spectroscopy does provide some insights on this complex
field.  
%
The velocity distribution of the $z = 0.36$ group members is
non-Gaussian, even with nearly 50 redshifts (McKean et al., in prep),
suggesting that the group is in a dynamically disturbed state.  
%
The velocity distribution in the B2108 field and the presence of
multiple elongated diffuse components connected by an apparent bridge
of emission suggest that system is undergoing some kind of merger.
Interestingly, the east-west alignment of the two diffuse X-ray
components is roughly in the same direction as 
%
the low surface brightness lobes seen in deep radio imaging of this
system \citep{more2108}, although the radio lobes cover a much smaller area
than the X-ray emission.

\subsection{Detection Rate and Group Luminosities}

One of the most basic quantities that emerges from the analysis of the
lens-group sample is the rate at which these lens-selected groups are
detected when observed at X-ray frequencies.  We have examined the
fields of six lenses, in which at least seven galaxy groups have been
discovered using optical spectroscopy.  Of those seven groups, four
are detected with the {\em Chandra} observations, giving a formal detection
rate of $\sim$60\% $\pm$30\%.  
All of the detected lens-selected groups have luminosities within
$R_{500}$ of greater than $10^{42}\ h^{-2}$ erg s$^{-1}$.
Because most previous investigations of the X-ray properties of galaxy
groups have been conducted at low redshifts, we compare the rate at
which X-ray emission from the IGM is detected in
the lens sample to these samples.  \citet{mahdavi00} selected groups
and clusters based on the CfA redshift survey and found a detection
rate of 23\%, which they correct to a 40\% rate based on their X-ray
selection function.  The X-ray data used for this work were from the
ROSAT All-Sky Survey \citep[RASS;][]{rass} and thus are not highly
sensitive, with a limiting luminosity of $\sim$10$^{42}$ erg
s$^{-1}$. \citet{gemsxray} concentrate more specifically on groups and
detect $\sim$60\%, based on deeper pointed X-ray observations
with ROSAT.  For this sample, the limiting luminosity is $L_X \sim 10^{41}$
erg s$^{-1}$.  However, analysis and sample-selection issues make a
straight comparison of detection rates in these earlier samples
somewhat problematic.  The shallow X-ray data used in
\citet{mahdavi00} make it difficult to disentangle IGM emission from
emission due to a hot halo associated with one of the group galaxies
\citep[e.g.,][]{mulchaey_araa,gemsxray}.  The
\citet{gemsxray} sample avoids this problem by using deeper ROSAT
pointings, but their sample may be biased.  Their groups are
optically-selected, but they require their groups to have pre-existing
deep ($t_{\rm exp} > $10,000 s) ROSAT data, which 
%
were often available because the RASS data had shown an X-ray
source at that location.

Perhaps the best comparison sample for our lens-selected groups is the
work on local optically-selected groups by \citet{rasmussen06}.  Here,
the selection is unbiased, at least with respect to the X-ray properties,
and the X-ray observations, which are made with XMM, are sensitive.
The IGM has been detected in two out of the four groups, giving a
similar detection rate to the lens-selected samples.  We note,
however, that none of the groups in this sample have luminosities
above $10^{42}\ h^{-2}$ erg s$^{-1}$.  
Although both the lens-group and \citet{rasmussen06} samples are 
small, we may be seeing a difference in the sample properties.
The optically-selected sample may be more representative of typical
low-redshift galaxy groups, which often have low levels of X-ray
emission.  In contrast, the lens-selected sample is picking out groups that
are more like the X-ray--bright groups detected in local samples,
which tend to be on the high-mass side of the local group distributions
\citep[e.g.,][]{mahdavi00,mz98}.

\subsection{$L_X$--$\sigma_v$ relationship}

In order to compare the properties of the X-ray emission from the
lens-selected moderate-redshift groups with other group samples, we
have assumed $h = 0.7$ and $(\Omega_M,\Omega_\Lambda) = (0.3,0.7)$.
The lens-group points are plotted as the large stars on the
$L_X$--$\sigma_v$ plot in Figure~\ref{fig_lx_sig}.  
%
For the purposes of this plot, we have assumed that all of the
measured redshifts in the B2108 field are associated with the B2108-1
group, and plotted the derived velocity dispersion against the B2108-1
X-ray luminosity.
%
Also included in
the plot are data from the low-redshift X-ray--detected groups in the
GEMS sample of \citet{gemsxray}, and from X-ray--selected
moderate-redshift group samples \citep{jeltema06,xmmlss_groups}.  The
filled squares represent data on the four groups from
\citet{rasmussen06}.  As mentioned above, these optically-selected
groups appear to be less X-ray luminous than typical systems in the
X-ray--selected samples (Fig.~\ref{fig_lx_sig}). In contrast, the
lens-selected groups appear to be consistent with the X-ray--selected
samples, given the scatter in the data.

\ifsubmode 
\clearpage
\begin{figure}
\else 
\begin{figure*} 
\fi 
\plotone{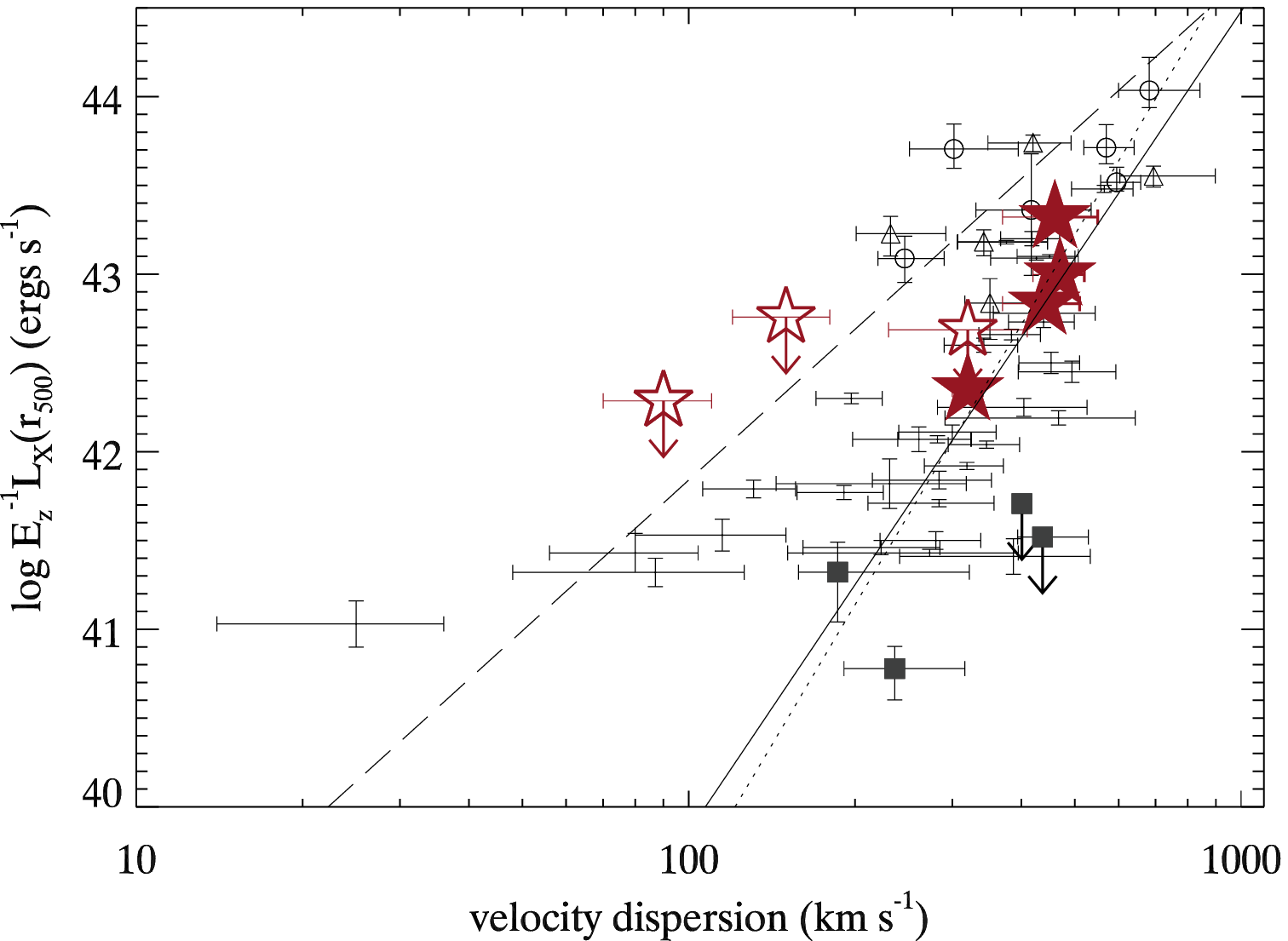}
\caption{
$L_X$--$\sigma_v$ plot showing moderate-redshift lens-selected groups
from this paper (large stars, with open stars representing upper
limits), low-redshift X-ray--selected groups from GEMS \citep[][error
bars with no points]{gemsxray}, moderate-redshift X-ray--selected
groups from \citet[][open circles]{jeltema06} and
\citet[][open triangles]{xmmlss_groups}, and low-redshift 
optically-selected groups from 2dF \citep[][filled
squares]{rasmussen06}.  The solid line represents the fit to the GEMS
sample \citep{gems_refit}, while the dotted line is the fit to
the low-redshift cluster data of \citet{markevitch98} and
\citet{horner_thesis}.  The long-dashed line is a fit to 
a sample of intermediate-redshift groups and clusters \citep{modz_lxsigfit}.
}
\label{fig_lx_sig}
\ifsubmode 
\end{figure}
\clearpage
\else 
\end{figure*} 
\fi 

Also shown in Fig.~\ref{fig_lx_sig} are fits to low- and
intermediate-redshift samples.  The low-redshift scaling relations are
shown for the GEMS group sample \citep[solid line;][]{gems_refit} and
a cluster sample \citep[short-dashed
line;][]{markevitch98,horner_thesis}.  The other scaling relation
(long-dashed line) in the plot is a fit made by \citet{modz_lxsigfit}
to data from an intermediate-redshift group sample, the majority of
which is comprised of the \citet{jeltema06} and \citet{xmmlss_groups}
data points represented as the open circles and triangles in
Fig.~\ref{fig_lx_sig}.  It is clear that the lens-group detections are
more consistent with the low-redshift scaling relations than the
intermediate-redshift relation.  However,  \citet{modz_lxsigfit} do
point out that their fitted slope may be flatter than it should be
because the velocity dispersion measurements may be biased low.  
Of course, it is difficult to draw strong conclusions from only
four detections in the lens-group sample, but the upper limits, while
not constraining the properties of the lens-selected sample in a
meaningful way, are still perfectly consistent with the low-redshift
scaling relations and marginally consistent with the
intermediate-redshift scaling relation.  It should be noted that the
three lens-selected groups that were not detected all lie at $z>0.4$,
while the four that were detected all have $z < 0.4$.  Deeper X-ray
observations of the B1608 and B1600 fields, which now each have only
$\sim$30~ks of {\em Chandra} data, would enhance the conclusions
that can be drawn from the lens group sample.

\subsection{BGG offsets}

In low-redshift samples, the brightest group galaxy (BGG) nearly
always sits at the spatial and dynamical center of the group 
\citep[e.g.,][]{zm98}, and the diffuse X-ray emission from
group is also centered on the BGG
\citep[e.g.,][]{mz98,helsdonponman00,gemsxray}.  The coincidence
between the diffuse X-ray gas and the BGG becomes less prominent in
samples of X-ray--selected moderate redshift groups, 
%
where some groups have negligible BGG offsets but others have
offsets up to $\sim160 h^{-1}$ kpc
%
\citep{jeltema06,jeltema07}.  The lens-selected systems presented in
this paper span a redshift range similar to the \citet{jeltema07}
sample and thus provide an interesting comparison sample.  We have
identified the brightest spectroscopically-confirmed member of each of
the groups for which diffuse X-ray emission has been detected, either
from our own imaging (for B0712 and B2108) or from the data in
\citet[][for PG1115 and B1422]{momcheva}.  
%
The X-ray centroids are derived from the Gaussian-smoothed images
in the last column of Figure~\ref{fig_smoothximg}.  For the B2108
field, we calculate the offset between the BGG and the centroid of
B2108-1.  
%
The offsets between the BGGs and the X-ray centroids are given in
Table~\ref{tab_offsets}, and overlays of the X-ray contours on HST
images are shown in Figure~\ref{fig_xray_opt_overlay}.  For B2108 the
BGG is also the lensing galaxy.  However, in B0712 (because the group
lies at a lower redshift than the lens), PG1115, and B1422 the lens is
not the BGG.  Therefore, Table~\ref{tab_offsets} also gives the
offsets between the lensing galaxies and the X-ray centroids.

\ifsubmode 
\clearpage
\begin{deluxetable}{lllrrllrr}
\else 
\begin{deluxetable*}{lllrrllrr} 
\fi 
\tabletypesize{\scriptsize}
\tablecolumns{7}
\tablewidth{0pc}
\tablecaption{Offsets from X-ray Centroid}
\tablehead{
\colhead{}
 & \colhead{BGG}
 & \colhead{BGG}
 & \colhead{Offset}
 & \colhead{Offset}
 & \colhead{Lens Galaxy}
 & \colhead{Lens Galaxy}
 & \colhead{Offset}
 & \colhead{Offset}
\\
\colhead{Lens System}
 & \colhead{RA}
 & \colhead{Dec}
 & \colhead{(arcsec)}
 & \colhead{($h^{-1}$ kpc)}
 & \colhead{RA}
 & \colhead{Dec}
 & \colhead{(arcsec)}
 & \colhead{($h^{-1}$ kpc)}
}
\startdata                    
B0712   & 07:16:05.01 & +47:09:04.8 & 16 & 49 & 07:16:03.58 & +47:08:50.0 &
  9\tablenotemark{a} & \nodata\tablenotemark{a} \\
PG1115  & 11:18:15.52 & +07:45:47.7 & 16 & 51 & 11:18:17.00 & +07:45:57.7 &
 10\phantom{$^a$} & 31 \\
B1422   & 14:24:38.39 & +22:55:53.5 &  7 & 24 & 14:24:38.09 & +22:56:00.6 &
  1\phantom{$^a$} &  3 \\
B2108-1 & 21:10:54.03 & +21:31:00.4 &  7 & 25 & 21:10:54.03 & +21:31:00.4 &
  7\phantom{$^a$} & 25 \\
\enddata
\label{tab_offsets}
\tablenotetext{a}{The lens is not physically associated with the group.}
\ifsubmode 
\end{deluxetable}
\clearpage
\else 
\end{deluxetable*} 
\fi 

\ifsubmode 
\clearpage
\begin{figure}
\else 
\begin{figure*} 
\fi 
\plotone{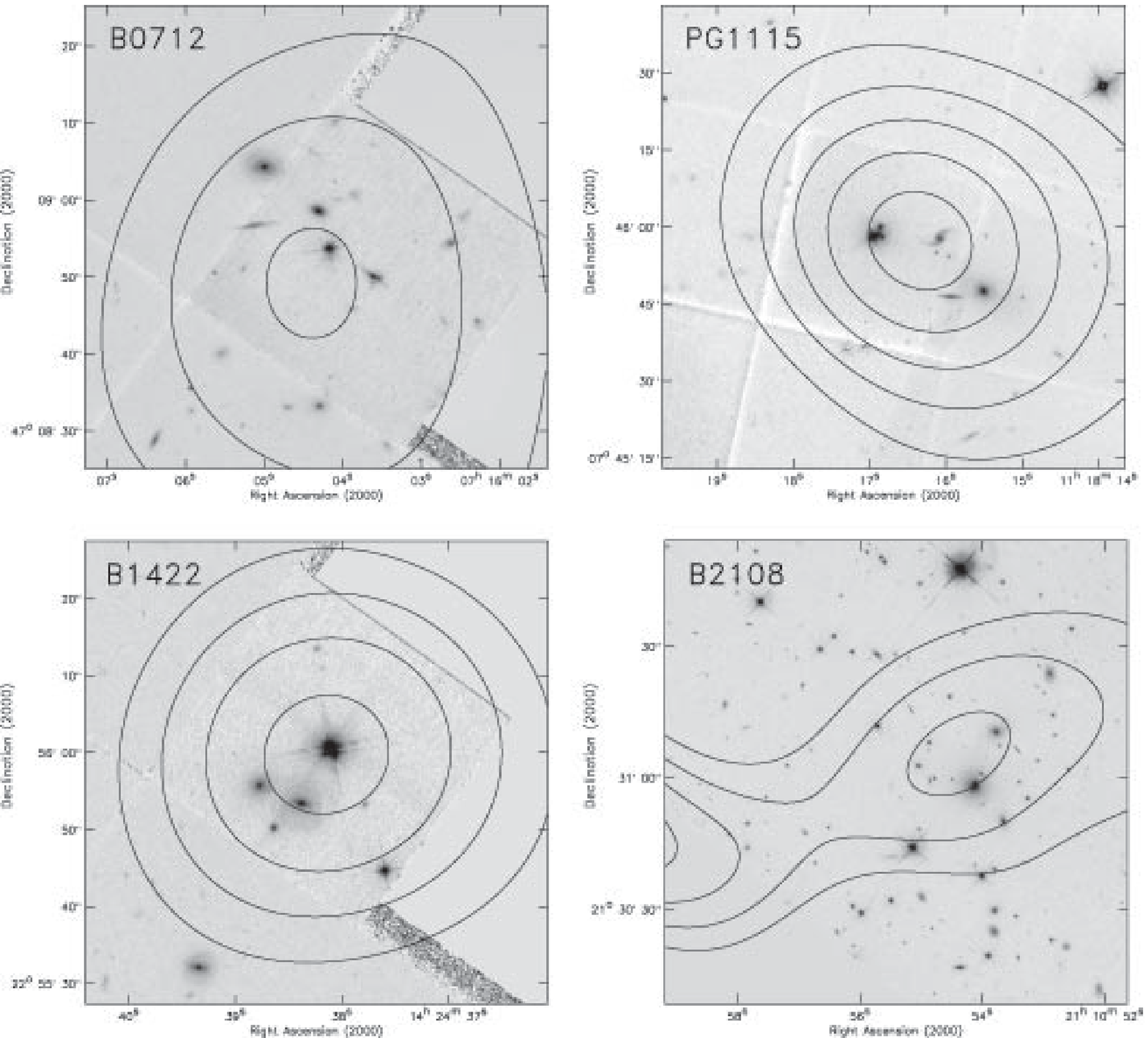}
\caption{
Optical imaging from the HST of the X-ray--detected groups, with contours
representing the diffuse X-ray emission overlaid.  For the B0712,
PG1115, and B2108 images, the data were taken with the F814W filter.
For the B1422 image, the data were taken with the F791W filter.
In each image the BGG is marked with a circle and the lens system is
marked with a box.  Note that the lens system in the B0712 group is
not physically associated with the group.
}
\label{fig_xray_opt_overlay}
\ifsubmode 
\end{figure}
\clearpage
\else 
\end{figure*} 
\fi 

The offsets between the centroids of the diffuse X-ray emission
and the BGGs are small to moderate, with all of the offsets being
$\leq$20\arcsec.  At the redshifts of these groups, $z \sim 0.3 -
0.4$, these offsets correspond to physical distances of
$\sim$25--50~$h^{-1}$~kpc.  For comparison, five out of the seven
groups in the sample of \citet{jeltema07} had BGG offsets of
$<$15~$h^{-1}$~kpc, while the other two had offsets of
110--160~$h^{-1}$~kpc.  
%
In spite of the BGG--X-ray offsets measured in the lens-group sample,
we cannot yet conclude that this sample is similar in character to the
\citet{jeltema07} one.  This is because slight changes in analysis
techniques can lead to significant changes in the centroid position of
lens-selected groups, as can be seen by comparing our X-ray images of
PG1115 and B1422 to those in \citet{grant_chandra}.  Even with the
excellent angular resolution of {\em Chandra}, the determination of
the morphology of faint diffuse emission in the presence of bright
lensed AGN images is challenging.  Over- or under-subtracting the AGN
emission can bias the centroid position.  
%
Furthermore, the choice of
smoothing technique can also lead to shifts in the derived X-ray
centroid.  
%
Deeper X-ray imaging of all of the systems would help to address
these centroiding issues.

Another complexity that enters into the determination of the
BGG--X-ray offsets is the difficulty for some of the systems in
identifying the BGG.  For many of the groups there is no clearly
dominant central galaxy.  Instead, in several of the groups (B0712,
B1608-3, B2108, and possibly PG1115) the brightest and next brightest
galaxies are separated by less than half a magnitude.  Other
investigations of intermediate-redshift groups
\citep{mulchaey06,jeltema06} have found systems for which the IGM is
detected in X-rays, but no dominant BGG was identified.  For our
lens-group sample, the lack of a dominant BGG in some groups may be
due to seeing these groups at an earlier stage of their evolution
\citep[e.g.,][]{jeltema07}, or it may just be due to incomplete
spectral information where the BGG has not yet been identified as a
group member.  Deeper optical data are needed to resolve this issue.

\subsection{Implications for Gravitational Lensing}

When a gravitational lens system resides in an overdense environment,
that environment can impact the lensing potential and have clear
effect on cosmological and astrophysics measurements made using the
lens \citep[e.g.,][]{keeton_group}.  An advantage of X-ray
observations over spectroscopic surveys of the lens environments is
that, in theory, the X-ray data can produce clear determinations of the
center of the group potential and the total group mass (through the
temperature of the X-ray gas).  However, the existing {\em Chandra} data
presented here are not sensitive enough to make robust measurements
of either the centroids or the temperatures of the X-ray gas distributions.

\subsection{Galaxy Properties}

In low-redshift groups, correlations have been found between the
morphological distribution of the member galaxies and the overall
group properties \citep[e.g.,][]{zm98,mz98,gemsxray}. In particular,
the fraction of early-type galaxies shows a significant correlation
with such properties as the group velocity dispersion and whether the
group has detectable diffuse X-ray emission.  At the redshifts of the
lens-selected sample, it is not possible to use ground-based imaging
to robustly determine galaxy morphologies.  Therefore, we have used
HST imaging to determine morphological properties of the group
members, where possible.  For nearly all of the systems, we used
archival WFPC2 or ACS imaging (GO-5699, PI: Impey; GO-5908, PI:
Jackson; GO-6555, PI: Schechter; GO-6652, PI: Impey; GO-7495, PI:
Falco; GO-8628, PI: Impey; GO-9133, PI: Falco; GO-9138, PI: Impey;
GO-9744, PI: Kochanek).  Most of these archival data sets were
obtained by the CfA-Arizona Space Telescope Lens Survey team.  For
B1608 we used our own deep ACS imaging (GO-10158: PI:
Fassnacht). Table~\ref{tab_f_e} lists the proportion of galaxies
identified by morphology as early-type (E or S0) in the lens-selected
groups. All of the groups for which diffuse X-ray emission has been
detected have relatively high early-type fractions, ranging from
$\sim$50--70\%.  For galaxy groups in the local Universe, the groups
for which a diffuse intragroup X-ray component is detected also have
significant early-type fractions.  In contrast, the group associated
with B1600, which has no detected X-ray emission, has only one
early-type galaxy among its confirmed members.  The B1608 groups are
mixed, with B1608-1 having a low early-type fraction and B1608-3
having a high fraction.  It may be that diffuse X-ray emission was not
detected in the B1608-3 group simply because the group is at a high redshift
and the observations were not particularly deep.  It should be noted
that the results in Table~\ref{tab_f_e} must be considered with
caution.  First, the number of member galaxies that have
high-resolution imaging is small because of the fields of view of the
HST cameras are limited.  Secondly, the fractions may well be biased
because the small fields are centered on the lens galaxy and may be
probing special regions of the groups, such as the group centers.

\ifsubmode 
\clearpage
\fi 
\begin{deluxetable}{lrrrl}
\tabletypesize{\scriptsize}
\tablecolumns{5}
\tablewidth{0pc}
\tablecaption{Group Early-type Fractions}
\tablehead{
 \colhead{Lens System}
 & \colhead{$N_{\rm tot}$}
 & \colhead{$N_{HST}$}
 & \colhead{$N_{\rm E/S0}$}
 & \colhead{$f_e$}
}
\startdata                    
B0712   & 15 &  9 &  5 & 0.56 \\
PG1115  & 13 &  7 &  5 & 0.71 \\
B1422   & 16 &  7 &  5 & 0.71 \\
B1600   &  7 &  7 &  1 & 0.14 \\
B1608-1 & 10 &  9 &  2 & 0.22 \\
B1608-3 &  8 &  7 &  6 & 0.85 \\
B2108   & 47 & 24 & 15 & 0.62 \\
\enddata
\label{tab_f_e}
\end{deluxetable}
\ifsubmode 
\clearpage
\fi 

\subsection{The Nature of the Lens-selected Groups}

Although the lens-selected group sample is small, the trends from the
sample are suggestive.  As discussed in the previous sections, the systems
for which diffuse X-ray emission has been detected appear to be
quite similar to local X-ray--loud samples in their luminosities,
their locations on the $L_X$--$\sigma$ plot, and their early-type
fractions.  This may not be totally unexpected, as the lensing galaxies
in strong lens systems tend to have early-type morphologies, and
local groups containing massive ellipticals are more likely to have 
detectable X-ray emission \citep[e.g.,][]{mulchaey96,mz98}.
In fact, the one group in the lens-selected sample in which the lensing
galaxy is clearly a spiral also has the lowest early-type fraction and
is one of the X-ray non-detections.  Thus, while the B1600
group and possibly the B1608-1 group appear to be like the typical group
found in the local Universe, with low masses and low early-type fractions
\citep[e.g.][]{gellerhuchra,eke04}, the majority of the lens-group
sample is more similar to the local groups for which X-ray emission
from the IGM has been detected, which are more massive and dominated
by early-type galaxies.  The one lens group that does not fit this
picture is the B1608-3 group, with its high early-type fraction and a
relatively high velocity dispersion.  The non-detection of B1608-3 in
the {\em Chandra} observations may be due to a combination of its
redshift and the short exposure time used in the observations.  We
feel that this is the group most likely to be detected if deeper
observations of the field are undertaken.

\section{Summary and Future Work}

In this paper we have presented the results of a systematic analysis
of ground-based optical spectroscopy and {\em Chandra} observations of
fields containing both strong gravitational lenses and previously
known moderate-redshift galaxy groups.  Diffuse emission from four of
the seven groups were detected in the X-ray observations, and the
properties of these four groups were found to be similar to
low-redshift X-ray--detected groups.  For these four lens systems, we
have found associated groups that appear to be gravitationally bound
structures, based on both the detection of diffuse X-ray emission from
the IGM and the redshift distribution of the group members.  Although
these are small-number statistics, it appears that galaxy groups
selected by their association with gravitational lenses are massive
and X-ray luminous.  Therefore, it may be possible to find massive
groups at moderate redshifts by conducting targeted and sensitive
X-ray observations of lens systems.

We note that most of the lens-selected groups were detected only at
low significance and thus a full analysis of their X-ray properties
was not possible.  For example, precise X-ray temperatures could not be
determined for any of the groups, and the determinations of
the group centroids also remain highly uncertain.  Therefore, deeper
X-ray observations of the sample, in particular of the B1600 and B1608
fields, are important to clarify the conclusions that can be drawn
from this non-traditional sample of moderate-redshift groups.
Similarly, more extensive optical spectroscopy of the lens fields
would allow better determinations of group memberships, velocity
dispersions, and the identities of the BGGs.  Furthermore, additional
lens-group associations have been identified spectroscopically and are
at redshifts that can be probed by deep {\em Chandra} observations.
Having a larger sample is key to quantifying the properties of
lens-selected galaxy groups.

\acknowledgments

We thank the anonymous referee for suggestions that improved the
paper.  
Support for this work was provided by the National Aeronautics and
Space Administration through Chandra Award Numbers GO3-4167X and
GO6-7125X issued by the Chandra X-ray Observatory Center, which is
operated by the Smithsonian Astrophysical Observatory for and on
behalf of the National Aeronautics Space Administration under contract
NAS8-03060.
This work is based in part on observations made with the National
Aeronautics and Space Administration (NASA)/ESA Hubble Space
Telescope, obtained from the Data Archive at the Space Telescope
Science Institute (STScI). STScI is operated by the Association of
Universities for Research in Astronomy, Inc., under NASA contract
NAS5-26555. These observations are associated with program AR-10300,
supported by NASA through a grant from STScI.
This work is supported in part by the European Community's Sixth
Framework Marie Curie Research Training Network Programme, Contract
No.  MRTN-CT-2004-505183 `ANGLES'.
Some of the data presented herein were obtained at the W. M. Keck
Observatory, which is operated as a scientific partnership among the
California Institute of Technology, the University of California, and
the National Aeronautics and Space Administration. The Observatory was
made possible by the generous financial support of the W.M. Keck
Foundation.  The authors wish to recognize and acknowledge the very
significant cultural role and reverence that the summit of Mauna Kea
has always had within the indigenous Hawaiian community.  We are most
fortunate to have the opportunity to conduct observations from this
mountain.
We are thankful for the dedication and expertise of the staffs of
{\em Chandra}, HST, and the Keck observatories, without whom
these observations would not have been possible.



\ifsubmode 
   {\it Facility:} \facility{CXO ()}, \facility{HST (ACS,WFPC2)}, \facility{Keck:I (LRIS)}, 
\fi 


\appendix

\section{Serendipitous Sources}

In addition to studying the B0712 and B2108 groups, we also searched
for diffuse and point-like serendipitous sources in the fields of each
system.  Object detection was carried out with the CIAO tool {\tt
wavdetect} \citep{wavdetect}, which was run with a detection
threshold of $10^{-6}$ on the ACIS chips 2,3,5,6,7.  We used scales
following the standard $(\sqrt2)^{i}$ series with a maximum wavelet
radius of 32 pixels.  Sources were detected in both the soft (0.5-2.0
keV) and hard (2.0-8.0 keV) bands separately and cross-correlated;
those detected with a $3\sigma$ significance or greater in at least
one band were included in the final source list.  Photometry was
carried out in the soft, hard and full (0.5-8.0 keV) bands for
point-like objects on the unfiltered event files using the {\tt ACIS
Extract} package.  Two diffuse sources were detected, one on chip 3
(ACIS-I3) of the B0712 dataset and one on chip 7 (ACIS-S3) of the
B2108 dataset; photometry of these sources were carried out manually
using the {\tt dmextract} task. The final source lists, including both
point-like and diffuse objects, along with the counts detected in all
three bands and the significances with which the sources were detected
are listed in Tables \ref{serendips_B0712} and \ref{serendips_B2108}.
The significances are calculated as the net counts from each source
divided by the error of the background counts in the measurement
aperture, $\sigma = {\rm NET\_COUNTS} / (1 + \sqrt{{\rm BKG\_COUNTS} +
0.75})$.

\ifsubmode 
\clearpage
\fi 
\begin{figure}[t]
\epsscale{1.0}
\plotone{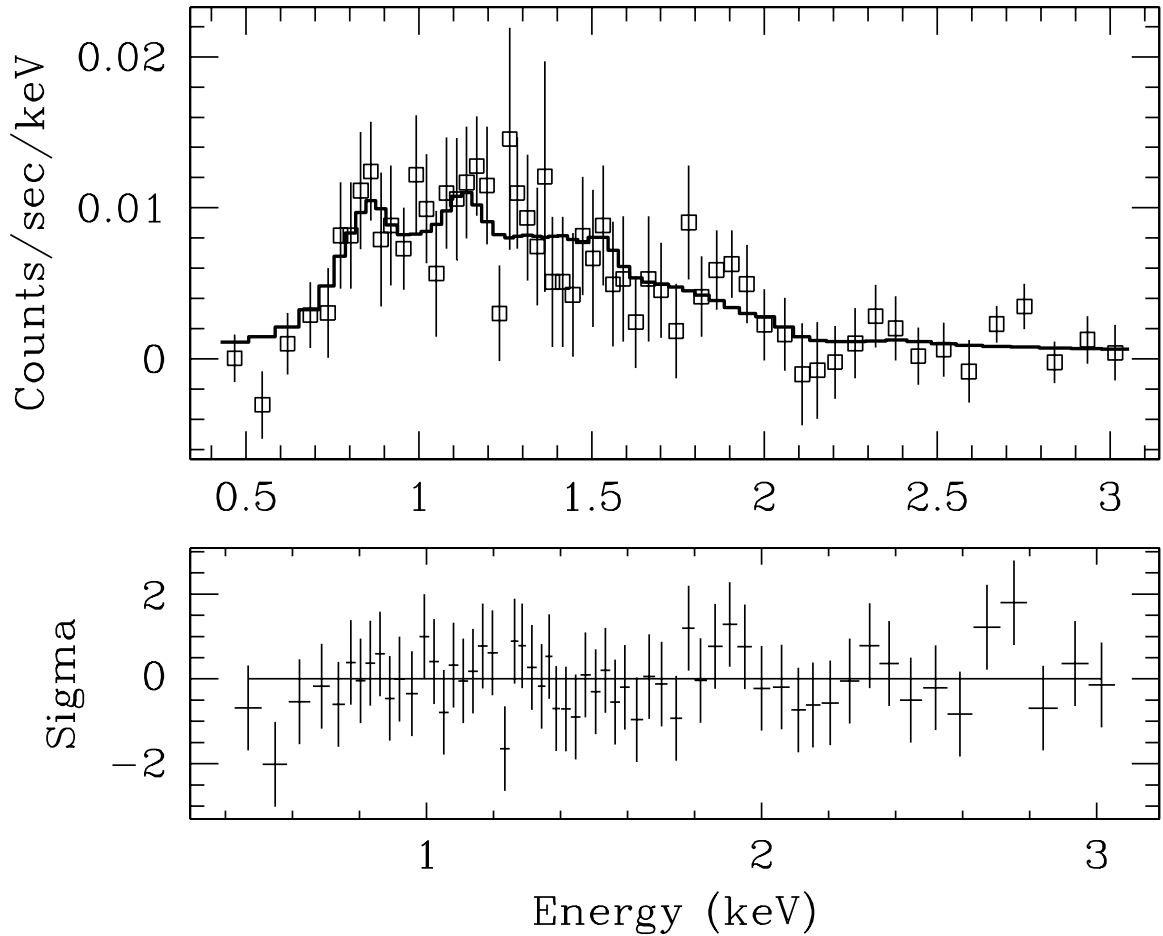}
\caption{
Spectral fit results to the observed energy spectrum of ZwCl
0713.1+4717.  The spectrum has been background subtracted and grouped
to contain at least 30 counts per bin.  The best fit Raymond-Smith
thermal spectrum has a temperature of 1.78 keV and a redshift of
0.30.}
\label{zwcl}
\end{figure}

\ifsubmode 
\clearpage
\fi 

\begin{deluxetable}{lccrrrrrr}
\tablewidth{0pt}
\tabletypesize{\scriptsize}
\tablecaption{Serendipitous Sources Detected in the Field of B0712+472}
\tablecolumns{9}
\tablehead{ 
 \colhead{Name}
 & \colhead{R.A.} 
 & \colhead{Dec}    
 & \colhead{Net Counts}               
 & \colhead{Net Counts}            
 & \colhead{Net Counts}               
 & \colhead{Sig} 
 & \colhead{Sig} 
 & \colhead{Sig}
 \\
 \colhead{}    
 & \colhead{(deg)}
 & \colhead{(deg)}  
 & \colhead{(Soft\tablenotemark{a})}  
 & \colhead{(Hard\tablenotemark{b})}
 & \colhead{(Full\tablenotemark{c})}
 & \colhead{(Soft\tablenotemark{a})}  
 & \colhead{(Hard\tablenotemark{b})}
 & \colhead{(Full\tablenotemark{c})}
}
\startdata
CXO J071654.3+471116\tablenotemark{d} & 109.22606 & 47.18790 & 1097 & 95 & 1192 & 37.6 & 2.0 &  21.3 \\
CXO J071628.8+472120 & 109.12033 & 47.35569 &  37 &  24 &  61 &  4.7 &  2.8 &  5.5  \\
CXO J071644.9+471858 & 109.18733 & 47.31631 &  15 &  26 &  41 &  2.6 &  3.3 &  4.4  \\
CXO J071629.6+471506 & 109.12341 & 47.25172 &  75 &  44 & 119 &  7.7 &  5.6 &  9.8  \\
CXO J071628.5+471725 & 109.11887 & 47.29028 &  25 &  30 &  55 &  3.7 &  3.8 &  5.6  \\
CXO J071643.4+472123 & 109.18121 & 47.35664 &  38 &  18 &  56 &  4.7 &  2.5 &  5.4  \\
CXO J071729.4+471445 & 109.37275 & 47.24597 &  64 &  43 & 106 &  6.2 &  4.3 &  7.7  \\
CXO J071713.6+471150 & 109.30704 & 47.19747 &  29 &  25 &  54 &  4.1 &  2.9 &  5.0  \\
CXO J071709.5+471259 & 109.28987 & 47.21639 & 134 &  56 & 190 & 10.2 &  5.4 & 11.6  \\
CXO J071641.8+471516 & 109.17429 & 47.25447 &  19 &  24 &  43 &  3.2 &  3.5 &  5.0  \\
CXO J071517.9+471713 & 108.82471 & 47.28706 & 109 &  29 & 138 &  8.9 &  2.5 &  8.4  \\
CXO J071516.4+472102 & 108.81854 & 47.35075 &  49 &   1 &  50 &  4.9 &  0.1 &  3.6  \\
CXO J071515.9+471614 & 108.81650 & 47.27078 &  21 &  20 &  41 &  3.3 &  2.4 &  4.1  \\
CXO J071510.9+471549 & 108.79575 & 47.26372 &  54 &  13 &  67 &  5.6 &  1.2 &  4.7  \\
CXO J071500.8+471819 & 108.75371 & 47.30533 &  34 &   0 &  31 &  3.8 &  0.0 &  2.3  \\
CXO J071459.0+472048 & 108.74605 & 47.34692 &  51 &   0 &  29 &  4.9 &  0.0 &  2.0  \\
CXO J071448.2+472044 & 108.70087 & 47.34566 &  32 &   0 &  26 &  3.6 &  0.0 &  2.0  \\
CXO J071442.8+472204 & 108.67871 & 47.36786 &  40 &  17 &  57 &  3.9 &  1.6 &  4.0  \\
CXO J071457.6+471628 & 108.74025 & 47.27464 &  26 &  26 &  52 &  3.2 &  2.1 &  3.6  \\
CXO J071545.1+471836 & 108.93796 & 47.31000 &  53 &  35 &  87 &  6.0 &  4.0 &  7.4  \\
CXO J071540.9+470852 & 108.92054 & 47.14789 & 122 & 210 & 332 & 10.1 & 13.5 & 17.2  \\
CXO J071558.5+471517 & 108.99392 & 47.25486 &  37 &  16 &  53 &  5.2 &  3.0 &  6.2  \\
CXO J071549.6+471143 & 108.95696 & 47.19545 &  18 &  16 &  34 &  3.4 &  3.1 &  4.9  \\
CXO J071559.9+471224 & 108.99966 & 47.20678 &  29 &  17 &  45 &  4.4 &  3.2 &  5.8  \\
CXO J071546.8+471404 & 108.94525 & 47.23461 &  70 &  44 & 113 &  7.4 &  5.6 &  9.6  \\
CXO J071533.7+471038 & 108.89050 & 47.17739 &  24 &  10 &  33 &  3.9 &  2.2 &  4.8  \\
CXO J071528.9+471729 & 108.87075 & 47.29144 &  45 &   0 &  41 &  5.4 &  0.0 &  4.3 \\
CXO J071525.1+471059 & 108.85475 & 47.18330 &  32 &   0 &  32 &  4.7 &  0.1 &  4.7 \\
CXO J071524.9+471104 & 108.85379 & 47.18467 &  37 &   2 &  39 &  5.1 &  0.7 &  5.2 \\
CXO J071516.3+471604 & 108.81796 & 47.26778 &  30 &  19 &  48 &  4.2 &  2.7 &  5.1  \\
CXO J071511.8+471451 & 108.79933 & 47.24767 &  26 &   9 &  35 &  3.7 &  1.4 &  3.9 \\
CXO J071550.1+471424 & 108.95908 & 47.24011 &  15 &   0 &  15 &  3.0 &  0.0 &  2.9 \\
CXO J071548.0+471709 & 108.95016 & 47.28608 &  19 &   1 &  19 &  3.3 &  0.2 &  3.1 \\
CXO J071545.8+471340 & 108.94100 & 47.22786 &  23 &  10 &  32 &  3.8 &  2.2 &  4.7  \\
CXO J071542.2+471333 & 108.92592 & 47.22603 &   1 &  18 &  19 &  0.5 &  3.2 &  3.3 \\
CXO J071532.8+470926 & 108.88704 & 47.15722 &   3 &  16 &  19 &  0.9 &  3.1 &  3.4 \\
CXO J071521.9+471141 & 108.84158 & 47.19483 &   0 &  22 &  22 &  0.0 &  3.8 &  3.8 \\
CXO J071522.1+471136 & 108.84217 & 47.19333 &   3 &  42 &  44 &  0.9 &  5.5 &  5.7 \\
CXO J071607.9+471428 & 109.03320 & 47.24111 &   7 &  16 &  22 &  1.7 &  3.0 &  3.7 \\
CXO J071610.7+470515 & 109.04496 & 47.08764 &  96 &  22 & 119 &  8.8 &  3.8 &  9.9  \\
CXO J071611.0+470856 & 109.04601 & 47.14903 &  26 &   8 &  33 &  4.2 &  1.9 &  4.8 \\
CXO J071620.3+471119 & 109.08471 & 47.18878 &  10 &  25 &  35 &  2.3 &  4.1 &  4.9  \\
CXO J071607.8+470613 & 109.03267 & 47.10375 &  23 &  20 &  43 &  3.8 &  3.5 &  5.5  \\
CXO J071627.6+470639 & 109.11521 & 47.11089 & 348 &  66 & 415 & 17.7 &  7.1 & 19.3  \\
CXO J071628.5+470508 & 109.11913 & 47.08558 &  21 &  37 &  58 &  3.6 &  4.9 &  6.4  \\
CXO J071636.6+470917 & 109.15279 & 47.15475 &  29 &   9 &  38 &  4.4 &  2.0 &  5.0 \\
CXO J071630.4+470525 & 109.12688 & 47.09031 &  25 &   0 &  24 &  4.0 &  0.0 &  3.7 \\
CXO J071625.9+470751 & 109.10829 & 47.13094 &  84 &   7 &  92 &  8.2 &  1.8 &  8.6 \\
CXO J071615.4+470323 & 109.06421 & 47.05661 &  54 &  14 &  68 &  6.3 &  2.5 &  7.0  \\
CXO J071610.6+471241 & 109.04441 & 47.21144 & 147 &   0 & 147 & 11.1 &  0.2 & 11.1 \\
CXO J071604.2+470751 & 109.01784 & 47.13097 &  22 &   5 &  26 &  3.8 &  1.3 &  4.2 \\
\enddata
\label{serendips_B0712}
\tablenotetext{a}{0.5-2.0 keV}
\tablenotetext{b}{2.0-8.0 keV}
\tablenotetext{c}{0.5-8.0 keV}
\tablenotetext{d}{Extended source}
\end{deluxetable}
\ifsubmode 
\clearpage
\fi 

We searched for known counterparts to the two detected diffuse sources
and found one to be 1.1 arcmin from the reported position of the
Zwicky cluster Zwcl 0713.1+4717 \citep{zwicky}.  We extracted the
spectrum of the cluster using the {\tt specextract} task and fit to it
an absorbed Raymond-Smith thermal plasma model with a 0.3 solar metal
abundance.  Allowing both the temperature and the redshift to vary,
our best-fit model returned a temperature of 1.78 keV and a redshift
of 0.30.  The extracted spectrum and our best-fit thermal model are
shown in Figure \ref{zwcl}.  We found no such counterpart to the
diffuse source detected in the field of B2108.  Furthermore, an
extraction of its spectrum proved inconclusive, as we could not fit to
it a thermal or power law model.

\ifsubmode 
\clearpage
\fi 
\begin{deluxetable}{lccrrrrrr}
\tablewidth{0pt}
\tabletypesize{\scriptsize}
\tablecaption{Serendipitous Sources Detected in the Field of B2108+213}
\tablecolumns{9}
\tablehead{ 
 \colhead{Name}
 & \colhead{R.A.} 
 & \colhead{Dec}    
 & \colhead{Net Counts}               
 & \colhead{Net Counts}             
 & \colhead{Net Counts}               
 & \colhead{Sig} 
 & \colhead{Sig} 
 & \colhead{Sig} 
\\
 \colhead{}    
 & \colhead{(deg)}
 & \colhead{(deg)}  
 & \colhead{(Soft\tablenotemark{a})}  
 & \colhead{(Hard\tablenotemark{b})}
 & \colhead{(Full\tablenotemark{c})}
 & \colhead{(Soft\tablenotemark{a})}  
 & \colhead{(Hard\tablenotemark{b})}
 & \colhead{(Full\tablenotemark{c})}
}
\startdata
CXO J211043.4+213404\tablenotemark{d} & 317.68085 & 21.56785 & 330 & 153 & 483 &  17.6 & 6.0 &  13.5 \\
CXO J211034.4+213837 & 317.64349 & 21.64375 &  13 &  23 &  36 &  2.6 &  3.4 &  4.5  \\
CXO J211022.2+214207 & 317.59253 & 21.70203 &  27 &   0 &  23 &  3.8 &  0.0 &  2.6  \\
CXO J211018.4+213708 & 317.57684 & 21.61889 &  18 &  11 &  28 &  3.1 &  2.0 &  3.9  \\
CXO J211007.6+214323 & 317.53174 & 21.72319 &  43 &  19 &  62 &  4.7 &  2.3 &  5.3  \\
CXO J211004.8+214358 & 317.52005 & 21.73281 &  23 &   0 &  17 &  3.1 &  0.0 &  1.7  \\
CXO J211116.8+213951 & 317.82001 & 21.66442 &  19 &  21 &  40 &  3.0 &  2.9 &  4.4  \\
CXO J211111.1+214013 & 317.79663 & 21.67033 &  74 &  54 & 128 &  7.4 &  5.9 &  9.7  \\
CXO J211055.3+214037 & 317.73053 & 21.67711 & 102 &  12 & 114 &  9.0 &  1.8 &  9.0  \\
CXO J211053.2+213831 & 317.72202 & 21.64206 &  19 &  11 &  29 &  3.3 &  2.2 &  4.2  \\
CXO J211048.7+214246 & 317.70322 & 21.71297 &  36 &  22 &  58 &  4.5 &  2.8 &  5.4  \\
CXO J211004.7+212510 & 317.51959 & 21.41945 & 222 &  17 & 238 & 13.3 &  1.6 & 12.5  \\
CXO J211003.5+212622 & 317.51474 & 21.43947 &  47 &  14 &  61 &  5.1 &  1.4 &  4.6  \\
CXO J211001.7+213149 & 317.50708 & 21.53047 & 105 &   5 & 109 &  8.4 &  0.5 &  7.0  \\
CXO J210959.5+212826 & 317.49814 & 21.47389 &  42 &  20 &  62 &  4.7 &  1.9 &  4.7  \\
CXO J210956.8+213023 & 317.48691 & 21.50642 &  34 &   0 &  30 &  3.9 &  0.0 &  2.4  \\
CXO J211034.2+213312 & 317.64288 & 21.55358 &  45 &  22 &  67 &  5.8 &  3.8 &  7.2  \\
CXO J211023.6+213310 & 317.59839 & 21.55289 &   4 &  22 &  27 &  1.3 &  3.7 &  4.1  \\
CXO J211020.8+213255 & 317.58698 & 21.54883 &  24 &  18 &  42 &  3.9 &  3.2 &  5.4  \\
CXO J211018.6+212955 & 317.57776 & 21.49875 &  24 &   1 &  25 &  3.9 &  0.4 &  3.9  \\
CXO J211013.0+212549 & 317.55420 & 21.43031 &  87 &  34 & 121 &  8.0 &  4.2 &  9.3  \\
CXO J211116.4+213250 & 317.81842 & 21.54736 & 374 & 207 & 581 & 18.3 & 13.3 & 23.0  \\
CXO J211114.2+213554 & 317.80923 & 21.59853 &  71 &  28 &  99 &  7.4 &  4.2 &  8.8  \\
CXO J211113.1+213443 & 317.80475 & 21.57875 &  16 &   5 &  21 &  3.0 &  1.1 &  3.3  \\
CXO J211108.6+213052 & 317.78604 & 21.51455 &  18 &   1 &  19 &  3.4 &  0.3 &  3.4  \\
CXO J211105.4+212950 & 317.77249 & 21.49747 &  22 &  10 &  32 &  3.7 &  2.3 &  4.7  \\
CXO J211104.2+212747 & 317.76785 & 21.46331 &   3 &  28 &  31 &  1.0 &  4.3 &  4.6  \\
CXO J211102.0+213444 & 317.75867 & 21.57889 &  17 &   4 &  22 &  3.3 &  1.2 &  3.7  \\
CXO J211056.7+213315 & 317.73660 & 21.55422 &  38 &  20 &  57 &  5.2 &  3.5 &  6.6  \\
CXO J211055.6+212732 & 317.73199 & 21.45903 &  20 &   4 &  24 &  3.5 &  1.2 &  3.9  \\
CXO J211052.6+213239 & 317.71918 & 21.54431 &  23 &  23 &  45 &  3.9 &  3.9 &  5.8  \\
CXO J211051.3+213459 & 317.71381 & 21.58311 &  16 &  16 &  32 &  3.1 &  3.1 &  4.6  \\
CXO J211050.1+213247 & 317.70898 & 21.54642 &  24 &   5 &  28 &  4.0 &  1.4 &  4.4  \\
CXO J211047.3+212804 & 317.69742 & 21.46795 &  70 &   1 &  70 &  7.4 &  0.2 &  7.4  \\
\enddata
\label{serendips_B2108}
\tablenotetext{a}{0.5-2.0 keV}
\tablenotetext{b}{2.0-8.0 keV}
\tablenotetext{c}{0.5-8.0 keV}
\tablenotetext{d}{Extended source}
\end{deluxetable}
\ifsubmode 
\clearpage
\fi 


\end{document}